\documentclass[12pt]{article}
\pdfoutput=1 

\usepackage{amssymb}
\usepackage{amsfonts}
\usepackage{amsmath}
\usepackage{graphicx}
\usepackage{enumitem} 

\usepackage{setspace} 

\usepackage{caption}
\usepackage{amssymb,amsmath,amsfonts,bbm}
\usepackage{graphicx}
\usepackage{mathrsfs}
\usepackage{calligra}
\usepackage{multirow}
\numberwithin{equation}{section}

\newcommand{\PMB}[1]{
\leavevmode
\setbox0=\hbox{#1}%
\kern-0.02em\copy0\kern-\wd0
\kern0.02em\copy0\kern-\wd0
\kern-0.025em\raise0.0167em\box0
\kern-0.025em\raise0.0433em\box0}

\begin{document}

\begin{center}
{\bf {\Large ODC and ROC curves, comparison curves, and stochastic dominance}{\Large {$^\star$}}}\\
\end{center}

\vspace{0.2cm}

\begin{center}
{\bf Teresa Ledwina}$^1$ and {\bf Adam Zagda\'nski}$^2$\\

\vspace{0.2cm}
$^1$ {\it Institute of Mathematics, Polish Academy of Sciences, ul. Kopernika 18, 51-617 Wroc{\l}aw, Poland, E-mail: ledwina@impan.pl}\\
$^2$ {\it Faculty of Pure and Applied Mathematics, Wroc{\l}aw University of Science and Technology, Wyb. Wyspia\'nskiego 27, 50-370 Wroc{\l}aw, Poland}

 \vspace{0.5cm}
\end{center}

\noindent

\begin{small}
\noindent
We discuss two  novel  approaches to inter-distributional comparisons in the classical two-sample problem. Our starting point are properly standardized and combined, very popular in several areas of statistics and data analysis, ordinal dominance and receiver characteristic curves, denoted by ODC and ROC, respectively. The proposed new curves are termed the comparison curves. Their estimates, being weighted rank processes  on (0,1), form the basis of inference. These  weighted processes are intuitive, well-suited for visual inspection of data at hand, and are also  useful for constructing some formal inferential procedures.  They can be applied to several variants of two-sample problem. Their use can help to improve some existing procedures both in terms of power and the ability to identify the sources of departures from the postulated model. To simplify interpretation of finite sample results we restrict attention to values of the processes on a finite grid of points. This results in the so-called bar plots (B-plots) which readably summarize the information contained in the data. What is more, we show that B-plots along with adjusted simultaneous acceptance regions provide principled information about where the model departs from the data. This leads to a framework which facilitates identification of regions with locally significant differences.

We show  an implementation of the considered techniques to a standard stochastic dominance testing problem. Some min-type statistics are introduced and investigated. A simulation study compares two tests pertinent to the comparison curves to well-established tests in the literature and demonstrates the strong and competitive performance of the former in many typical situations. Some real data applications illustrate simplicity and practical usefulness of the proposed approaches. A range of other  applications of considered weighted processes is briefly discussed too. \\
\end{small}

\noindent
{\Large {$^{\star}$ }}{{\sf This paper is dedicated to the memory of Professor Konrad Behnen (1941-2000).}}\\


\newpage
\begin{center}
{\bf Content}
\end{center}

\noindent
{\small 
\setlist{nolistsep} 
\begin{itemize}[noitemsep]
\item[1.] Introduction\dotfill\ 3 
\item[2.] Relative distributions\dotfill\ 5
	\begin{itemize}
	\item[2.1.]  Unpooled case: comparison curve CC related to ODC and ROC functions, \\ and its interpretation \dotfill\ 7 
	\item[2.2.] Pooled case: contrast comparison curve CCC related to two ODC  functions,\\ and its interpretation \dotfill\ 8
	\end{itemize}
\item[3.] Empirical variants of CC and CCC, related empirical processes, and B-plots\dotfill\ 9
\item[4.] Fourier coefficients of comparison densities and their empirical versions\dotfill\ 12
\item[5.] Testing for   stochastic dominance\dotfill\ 13
	\begin{itemize}
	\item[5.1.] Test statistics\dotfill\ 14
	\item[5.2.] Acceptance regions for subset of bars in B-plot\dotfill\ 15
	\item[5.3.] Simulation study \dotfill\ 16
		\begin{itemize}
		\item[5.3.1.] Alternatives\dotfill\ 17
		\item[5.3.2.] Competitors and empirical powers\dotfill\ 18
	\end{itemize}
	\item[5.4.] Real data examples\dotfill\ 20
		\begin{itemize}
		\item[5.4.1.] Income data analysis\dotfill\ 20
		\item[5.4.2.] Cholesterol data analysis\dotfill\ 23
		\end{itemize}
	\item[5.5.] Other testing problems on stochastic dominance or some related trends\dotfill\ 27
\end{itemize}
\item[6.] Conclusions\dotfill\ 27
\itemindent=-15pt 
\item[] Acknowledgements\dotfill\ 28
\item[] References\dotfill\ 28
\item[] Appendix A. On asymptotic distribution of bars\dotfill\ 32
\item[] Appendix B. Some results on $\;{\cal U}_N$\dotfill\ 35
\item[] Appendix C. Auxiliary simulation experiments\dotfill\ 38
\begin{itemize}
\item[C.1.] Canadian after-tax family income in 1978 versus 1986 comparison with D(N) = 16 383 \dotfill\ 39
\item[C.2.] Cholesterol level in obese men in Puerto Rico versus Honolulu  comparison with D(N) = 511\dotfill\ 40
\end{itemize}
\item[] Appendix D. Some comments on relation of CC and CCC\dotfill\ 42
\item[] Appendix E. New summary indices and related discussion\dotfill\ 43
\item[] References for Appendices\dotfill\ 44
\end{itemize}
}

\newpage

\noindent 
{\bf 1. Introduction}\\

\noindent
There is a vast amount of literature on comparing two samples. Comparing two populations or two treatments is a central question in statistical practice. In economic applications the aim may be to compare incomes or financial returns of two populations, for example. In medical studies nonparametric two-sample procedures are often designed to compare the responses of treatment and control groups, respectively. Specifically, many experiments are conducted to determine whether or not a new treatment is better than no treatment or other treatment. To simplify the discussion, we can imagine that the cumulative distribution function (CDF) $G$ describes the treatment effect while $F$ refers to the control group. For an illustration we consider a situation in which the treatment is expected to increase the response and this effect is formalized here as {\it the stochastic dominance}, or {\it the  stochastic order}, in other words. In general, we say that a distribution function $F$ is stochastically dominated by a distribution function $G$, if  $F(x) \geq G(x)$ for all $x \in \mathbb R$;  $F \geq G$, in short. 
Roughly speaking, the ordering means that the treatment tends to result in a larger response than the control, as a rule. The notion is not a restrictive one. Therefore, its invalidation indicates that many stronger notions are not adequate either.

In view of importance of two-sample comparisons several tests under several settings have been elaborated. In this paper we focus on continuous data on real line and nonparametric methods. Typical approach in this case is to consider certain distances  between empirical distribution functions, and to reject a null hypothesis when the selected distance is too large. Standard examples are Kolmogorov-Smirnov and Area Under Curve statistics, related to the supremum and $L_1$  distances, respectively. Parallel considerations concern distances of quantile functions. Additionally, more advanced approaches based for example on empirical likelihood, spacings, entropy, smooth tests etc. have been elaborated. For high-dimensional and more complex data some adjusted tools have been also worked up. 

However, a formal testing of a specified  null hypothesis solves only a part of the problem. In case of testing, except an obvious requirement to control the size of test and 
to have good power under reasonably large class of alternatives of interest, some other aspects of test construction and related decisions play a role, as well. In view of nowadays computer aided techniques many old problems  with the null distribution of test statistics can be solved. Hence, one can concentrate on another aspects which are important for practice. In particular, some evidence why the null model is invalidated is nowadays considered to be important to get. As nicely concluded by Cazals and Lh\'eritier (2015) ``...{\sf a rejection decision should be just a signal to examine further the data in order to understand}". Moreover, an insight into a structure of the data when null model is accepted is also welcome, since the acceptance can simply be due to some weaknesses of the applied procedure, which is not able to detect some types of departures.   Also, some visual inspection of data at hand is of interest, because capturing the key differences between the two populations in a parsimonious and easy-to-interpret  way is crucial for practice. For related discussion  in the two-sample and other settings see e.g. Doksum and Sievers
	(1976), Beach and Davidson (1983), Duong (2013), Rousselet et al. (2017), Goldman and Kaplan (2018), Kim et al. (2019), Ma and Mao (2019), Algeri (2021), 
	Arza et al. (2023), Xiang et al. (2023), and Ducharme and Ledwina (2022).

To answer such questions, it is important to understand an inherited structure of the two-sample problem on $\mathbb R$. For this purpose we shall start with recalling two types of so-called relative distributions. Their history and related terminology are briefly summarized in Section 2. As seen, the well known ODC and ROC curves are just basic objects in this area. Then, we introduce new tools, closely related to the ODC and ROC curves, called {\it the comparison curve} (CC) and {\it the contrast comparison curve} (CCC), respectively, which are better suited to detecting local differences between the two populations than the original curves. The material is followed by Section 3 where we introduce empirical variants of the comparison curves. These empirical curves define two natural weighted empirical rank processes on (0,1). Values of the processes over a selected  grid of points are termed as bar plots; B-plots in short. B-plots are our basis to detect both overall and local differences between $F$ and $G$. In Section 4 we discuss relation of B-plots to score vectors in some pertinent models in two samples. 
Section 5 concerns classical testing problem on stochastic dominance. We present two test
statistics and comment on their properties. Formal statements are postponed to Appendix
B. The simulation study reported there shows that new tests improve upon, standard in this
area, adjusted variants of Kolmogorov-Smirnov and AUC (area under the curve) procedures.
We also define acceptance regions for subsets of bars in B-plots, which provide clear and
principled auxiliary information about if, how, and where the two samples distributions differ. Next, we apply our methods to selected real data sets on income and cholesterol levels. We close with a brief discussion of other related testing problems.
Final conclusions are collected in Section 6. Five appendices provide additional information. 
In particular, Appendix A, based on existing literature,  presents some auxiliary technical results on joint distributions of bars. In Appendix B we formulate and prove some basic results on new test for stochastic dominance, defined in Section 5.1. Appendix C illustrates, by some simulation experiments, an influence of the size of the grid on the stability of the considered procedures. Appendix D contains some comments on relation of CC and CCC curves. In Appendix E we introduce and discuss new summary indices based on CC~and~CCC~curves.

\noindent
{\bf 2. Relative distributions} \\

\noindent 
Below, we recall two  ways in which two types of relative distributions have appeared in the literature. Their origins date back to the 1960s.

The first type of relative distribution steamed in nowadays literature from the receiver operating characteristic curve (ROC) methodology, originally developed for signal detection; cf. Green and Swets (1966) for an overview. Closely related approach, via the ordinal dominance curve (ODC), was introduced by Bamber (1975) and has stimulated further development. For a review of ROC methodology in the biomedical field see Pepe (2003) and Zhou et al.\ (2011). A general overview has been elaborated by Krzanowski and Hand (2009). The ROC curve is a simple graph summarizing the information contained in two population distributions. Nowadays, the list of available applications of this curve is impressive. For an insightful overview see Chapter 10 of Krzanowski and Hand (2009). On p.~15 of this book one can read: ``{\sf The ROC curve has a long history, and has been rediscovered many times by researches in different disciplines. This is perhaps indicative of its extreme usefulness, as well as of its naturalness\ldots}''.

To be specific, the first type of relative distribution of $G$ to $F$, on which we shall concentrate, is defined as $G(F^{-1}(p)),\; p \in [0,1]$, and is simply equal to the ODC$(p)$, when we adequately fix the axes for so-called sensitivity and specificity;  cf. Handcock and Morris (1999) for an extensive discussion. In the terminology of this book, $F$ is the {\it  reference distribution}. The ODC is simply the CDF of $F(Y)$, when $Y$ has the CDF $G$.
Huang and Pepe (2009) emphasize: ``{\sf This sort of standardization using a reference distribution is already commonplace in laboratory medicine and clinical medicine}''. In mathematical statistics ODC is known under the label the {\it  comparison distribution}; see Parzen (1998). Historically, under the name the {\it grade transformation}, the notion goes back to Galton. Presumably, the most popular name for $G(F^{-1}(\cdot))$ is PP {\it plot} of $G$ against $F$. Recently, Mathew (2023) has generalized the comparison distribution to {\it quantile cumulative distribution function} and discussed its applications in various fields of statistics. A number of nonparametric two-sample tests have been derived using the ODC function. Li et al. (1996) provides an impressive list of related references. For more recent impact see Carolan and Tebbs (2005), Davidov and Herman (2012), and Wang et al. (2020), for example.

In turn, the ROC curve, pertinent to $F$ and $G$, is defined by  ROC$(p)$=$1-G(F^{-1}(1-p))$ and formal arguments on both curves, ROC and ODC, are equivalent. ROC curves are most often considered in the context of classification and prediction problems; for an up-to-date presentation of these questions see Nakas et al. (2023).  A large stream of papers in ROC analysis  is concerned with  testing the equality or a comparison of two or more ROC curves. For an illustration see Wieand et al. (1989), Venkatraman (2000), Zhang et al. (2002), Zhang et al. (2018). For an  overview and several additional references see Fanjul-Hevia and Gonz\'alez-Manteiga (2018). In our contribution we employ and discuss ROC curve as a tool for inter-distributional comparison.
 
The second type of relative distribution, which we will discuss below, seems to have appeared in the mathematical statistics literature for the first time, somewhat in passing,  in the context of the asymptotic study of two-sample rank  process by Pyke and Shorack (1968), whose main concern was limit theory of linear rank statistics. The closely related idea of contrasting each individual empirical CDF with the respective pooled one can already be found in Kiefer (1959) and several related papers. However, latter on, such approach to testing has not attracted considerable attention for a long time. Formal definition of relative distribution of the second type can be found in Section 2 of Parzen (1977).  The notion has been rediscovered by Behnen (1981). The Behnen's paper introduces formally the respective relative distributions and uses them for a throughout discussion of inherited nature of nonparametric two-sample problem; see also Behnen and Neuhaus (1983, 1989) and related work for more easily accessible references and discussion. This, in turn, has allowed for substantial progress in modern test constructions for several variants of two-sample problem. For an illustration see Neuhaus (1987), Janic-Wr\'oblewska and Ledwina (2000), and Wy{\l}upek (2010), to concentrate on the most standard case.

To define the discussed  notion of relative distribution denote by $m$ and $n$ sample sizes from $F$ and $G$, respectively, and set $N=m+n$, $\;\lambda_N=m/N$, $H(x)=\lambda_NF(x)+(1-\lambda_N)G(x), \;x \in \mathbb R$. The CDF $H$ shall be the reference distribution in this approach. The relative distribution of $G$  to $H$ is  given by $G(H^{-1}(p)),\; p \in [0,1]$. Similarly one defines the relative distribution of $F$  to $H$. These relative distributions are mathematically very convenient.  Moreover, often, in view of experimental conditions,  they are  simply useful. For more details see Handcock and Morris (1999), Section 2.4.1. 

Our impression is that unpooled comparison distributions are more popular in many areas of applied statistics than pooled ones. Presumably a reason is that, at least at first glance, the  direct comparison of two CDF's $F$ and $G$ seems easier to interpret. However, we argue in Sections 2.1 and 2.2 that, by using the contrast comparison curve, proposed by us, results of both approaches  can be interpreted in a similar way. It even seems that the contrast comparison curve is easier to handle.

To simplify the description, in what follows we will rely on Parzen's terminology and call ODC(p) {\it the unpooled comparison distribution} and  refer to $G(H^{-1}(p))$ and to $F(H^{-1}(p))$ as {\it the pooled comparison distributions}.\\

\noindent
{\sf 2.1.  {Unpooled case: comparison curve} CC { related to} ODC { and} ROC { functions, and its interpretation}}\\[-1em]

\noindent
In this work, we propose to relate the inference to a normalized comparison distributions. In the unpooled case this results in a normalized ODC/ROC function, which we define as follows
$$
\text{CC}(p)=\frac{p-G\bigl(F^{-1}(p)\bigr)}{\sqrt{p(1-p)}}=\frac{p-\text{ODC}(p)}{\sqrt{p(1-p)}}=\frac{\text{ROC}(1-p)-(1-p)}{\sqrt{p(1-p)}},\;\;\;p \in (0,1).
\eqno(1)
$$ 
We shall call CC$(p),\;p \in (0,1),$ {\it the comparison curve}. It allows to inspect how allocation of probability mass related to $G$ differs from that pertinent to $F$, when compared at successive $p$-quantiles of $F$. 
CC$(p)$ is similar to the graph of  ODC$(p)$ plot in that it is invariant under strictly increasing and continuous  transformations of the measurement scale. On the other hand, due to the weight function in (1), CC$(p)$ much better exhibits differences in between $G$ and $F$ appearing in tails. In particular, it is quite often that CC$(p)$ is  unbounded for $p \approx 0$ or $p \approx 1$, while typically ODC$(p)$ is  0, as $p \to 0$, and  1, as $p \to 1$. The simplest example leading to unbounded CC$(p)$ is $F(x)=\Phi(x)$ -- the CDF of $N(0,1)$ distribution, and $G(x)=\Phi(x/\sigma),\;\sigma > \sqrt 2$.

For notational simplicity, we will mostly exploit the first expression in (1), with ODC$(p)=G(F^{-1}(p))$. Note that if $F(x)=G(x)$ for all $x \in \mathbb R$, then CC$(\cdot)= 0.$ Moreover, CC$(p) \geq 0$ for all $p \in (0,1)$ if and only if $F(x) \geq G(x)$ for all $x \in \mathbb R$, i.e. the stochastic dominance of $F$ by $G$ takes place. This is well known property; see e.g. Holmgren (1995). However, as noticed in Ducharme and Ledwina (2022), there is a  deeper connection between CC shape and stochastic dominance. Namely, if there is only one point $p=p_0, \;p_0 \in (0,1)$, such that CC$(p_0)=0$ then $F^{-1}(p_0)=G^{-1}(p_0)$ and, consequently, the set $(-\infty, F^{-1}(p_0)]$ has the same probability under $F$ and $G$. Obviously, similar conclusion  also holds for the complement of this set.
Note that the relation CC$(p) >0$ on $(0,p_0)$ defines the region where $F(\cdot) > G(\cdot)$. Hence, when restricted to this interval, the information CC$(p) >0$ means that 
observations generated from the pertinent conditional distribution of $G$ are stochastically larger than under respective conditional variant of $F(\cdot)$.  Otherwise stated, on this set, the mass associated with $G(\cdot)$ accumulates more intensively toward $F^{-1}(p_0)$ than the same amount of mass pertinent to $F(\cdot)$. The reverse holds when CC$(\cdot) < 0$ on $(0,p_0)$. If there are more than one point $p$ such that CC$(p)=0$, a similar interpretation applies to each of resulting regions of $p$'s.

With appropriate standardization (cf.\ (7) and Appendices A and B), changing magnitude of empirical variant of CC$(p)$ reflects in a convincing way the rate at which mass allocation between the two empirical CDF's changes.  
 
The above shows that using CC curve our understanding of the data and its structure can be improved. For additional material on this point see Section 5.2. 

Note that Li et al. (1996) have considered ODC$(p) - p,\;p \in (0,1),$ and called it the vertical shift function. In contrast to this development,  we do not only center ODC(p), but rescale it as well. This introduces adequate weights to each of the points $p,\;p \in (0,1)$. Recall also that Doksum (1974) has introduced the horizontal distance between $F(x)$ and $G(x)$, given by $\Delta(x)=G^{-1}(F(x))-x,$ for each $x \in \mathbb R$.

Related notion of {\it the comparison curve} in the context of goodness of fit testing has been recently introduced in Ducharme and Ledwina (2022). Hence, considerable amount of examples of CC's can be found there. We also provide some evidence in Section 5.3.1. A forerunner of the mentioned above comparison curves  is  {\it the quantile dependence function} introduced in Ledwina (2015) in the case  of independence testing.   \'Cmiel and Ledwina (2020) contains  graphs of this function for several known bivariate CDF's. Needless to say  that analogous comparison  curves can be introduced in many other testing problems.

To close this presentation we shall introduce also {\it the comparison density} related to the comparison distribution function $G\bigl(F^{-1}(p)\bigr)$. We shall use it to state and interpret some asymptotic results in Appendix A and in  discussion in Section 4.

Assume that both $F$ and $G$ are continuous with densities $f$ and $g$, respectively. Moreover, assume also that $G$ is absolutely continuous with respect to  $F$, what implies that $f(x)=0$ yields $g(x)=0$. Then the function ODC$(p)=G(F^{-1}(p))$ satisfies ODC(0)=0, ODC(1)=1, and possesses a density, called {\it the comparison density}, which is given by
$$
r(p)=\frac{dG(F^{-1}(p))}{dp}=\frac{g(F^{-1}(p))}{f(F^{-1}(p))},\;\;\;p\in (0,1).
\eqno(2)
$$
Detailed discussion on a history of the comparison distribution function and its density can be found in Handcock and Morris (1999), and Thas (2010). It should be noted that $r(p)$ is often unbounded. \\

\noindent
{\sf 2.2.  Pooled case: contrast comparison curve related to two ODC   functions, and its interpretation}\\[-1em]

\noindent
This approach is  inspired by work of Behnen (1981) and  Ledwina and Wy{\l}upek (2012a,b). Basing on these findings, a natural counterpart of CC$(p)$ has a form
$$
\text{CCC}(p)= \frac{F(H^{-1}(p))-G(H^{-1}(p))}{\sqrt{p(1-p)}}=\frac{(F-G)(H^{-1}(p))}{\sqrt{p(1-p)}},
\eqno(3)
$$
where $H(x)=\lambda_NF(x)+(1-\lambda_N)G(x),\;x \in \mathbb R,\; \lambda_N=m/N$, and shall be labeled as {\it the contrast comparison curve}, CCC in short. Since $H$ depends on $\lambda_N$ therefore the same concerns CCC$(p)$. However, we skip this in our notation.

The contrast comparison curve can be interpreted in the similar way as the comparison curve in Section 2.1. In particular,  $F = G$ if and only if CCC $=0$. The relation CCC$(p) \geq 0$ for all $p \in (0,1)$ is equivalent to $F(x) \geq G(x)$ for all $x \in \mathbb R$. If there is one point $p_0 \in (0,1)$ such that CCC$(p_0)=0$ and CCC$(p) > 0$ on $(0,p_0)$ then we know that restricting attention to $(-\infty, H^{-1}(p_0)]$, the  distribution of $G$ is stochastically larger than $F$, while the changing magnitude of CCC$(\cdot)$ reflects changing intensity of allocation of corresponding probability masses. The last term in (3) shows that, in fact, comparison of $F$ and $G$ via CCC is even more immediate than respective comparison via CC.

Similarly as in the unpooled case, we can relate the difference  $F(H^{-1}(p))-G(H^{-1}(p))$ to some comparison densities. Since, for any $\lambda_N \in (0,1)$, the distribution induced by $H$ dominates the distributions pertinent to $F$ and $G$, the densities introduced below always exists. Set
$$
c(p)=r_1(p)-r_2(p),\;\;\mbox{where}\;\; r_1(p)=\frac{d F(H^{-1}(p))}{dp},\;\; r_2(p)=
\frac{d G(H^{-1}(p))}{dp}.
\eqno(4)
$$ 
Again, it should be observed that $c(p), r_1(p)$ and $r_2(p)$ depend on $m$ and $n$ throughout $\lambda_N$. Note also that $\int_0^1c(p)dp=0$ and $\lambda_N r_1(p)+(1-\lambda_N) r_2(p)=1$. In consequence, the comparison densities $r_1(p)$ and $r_2(p)$,  as well as the contrast function $c(p)$ are always bounded. This is in sharp contrast to $r(p)$ introduced in the unpooled case. \\

\noindent
{\bf 3. Empirical variants of CC  and CCC,  related empirical processes, and B-plots}\\

\noindent
To introduce empirical variants of CC and CCC curves,  consider two samples $X_1,...,X_m$ and $Y_1,...,Y_n$ with continuous parent CDF's $F$ and $G$, respectively. Let $F_m$ and $G_n$ denote empirical distribution functions for the samples $X_1,...,X_m$ and $Y_1,...,Y_n$, respectively. As before,  $N=m+n\;$ and $\;\lambda_N=m/N$. We shall assume throughout that $0 < \lambda_* \leq \lambda_N \leq 1-\lambda_* < 1,$ for some $ \;\lambda_* \leq 1/2$.  Now introduce empirical variant $H_N$ of $H$, i.e.  $\;H_{N}(x)=\lambda_N F_m(x) + (1-\lambda_N) G_n(x)$,  $\;x \in \mathbb R$. Here $N$ stands for the pair $(m,n)$. Note that $H_N$ is an ordinary empirical CDF in the pooled sample.

With this notation,  empirical variant $\widehat{\text{CC}}(p)$ of CC$(p)$ takes the form
$$
\widehat{\text{CC}}(p)=\frac{p-G_n\bigl(F_m^{-1}(p)\bigr)}{\sqrt{p(1-p)}},\;\;\;p\in (0,1).
\eqno(5)
$$  
In turn, natural estimate $\widehat{\text{CCC}}(p)$ of CCC$(p)$ is given by
$$
\widehat{\text{CCC}}(p)=\frac{F_m(H_{N}^{-1}(p))-G_n(H_{N}^{-1}(p))}{\sqrt{p(1-p)}},\;\;\;p \in (0,1).
\eqno(6)
$$

\noindent
The estimators (5) and (6) are invariant to strictly increasing and continuous transformations of observations in the pooled sample $(X_1,...,X_m,Y_1,...,Y_n)$. This implies that, under $F = G$, $\widehat{\text{CC}}(p)$ and $\widehat{\text{CCC}}(p)$ are distribution free. Note also that,  under $F = G$ and arbitrary fixed $p \in (0,1)$, our estimates satisfy
$$
\eta_N \widehat{\text{CC}}(p) \stackrel{\cal D}{\longrightarrow} N(0,1),\;\;\;\mbox{and}\;\;\;\eta_N \widehat{\text{CCC}}(p) \stackrel{\cal D}{\longrightarrow} N(0,1),\;\; \eta_N=\sqrt{\frac{mn}{N}},
\eqno(7)
$$
cf. Appendix A. These simple relations are consequences of general results on rank empirical processes, which we quote therein. In this context, it is worth of noticing that useful paper by Aly et al. (1987) seems to  be overlooked in ROC literature.


As for the possible application  of (5) and (6) in testing, their forms suggest to proceed in a traditional way in mathematical statistics and to consider  some classical functionals of these weighted processes to get some test statistics for verifying different forms of relations of $F$ and $G$. Also, some confidence regions could be constructed. However, basing on our earlier experience and remembering tasks discussed in Section 1, we shall consider easier tractable objects, which provide sufficient information to build sensitive tests and to get simultaneously reliable insight into sources of possible invalidation of the model. The idea is to consider a grid of points and values of the weighted processes 
$$
{\sf U}_N(p)=\eta_N\frac{p-G_n(F_m^{-1}(p))}{\sqrt{p(1-p)}}=\eta_N \widehat{\text{CC}}(p),\;\;p \in (0,1),\;\;\;
\eqno(8)
$$
and
$$
{\sf P}_N(p)=\eta_N\frac{F_m(H_{N}^{-1}(p))-G_n(H_{N}^{-1}(p))}{\sqrt{p(1-p)}}= \eta_N \widehat{\text{CCC}}(p),\;\;p \in (0,1),
\eqno(9)
$$
over this grid. Our motivations are explained in more details in Sections 4 and 5.2. The grid, we shall use,  follows from from diadic partition of [0,1]. More precisely, given $s,\;s=0,1,...,$ set $d(s)=2^{s+1}-1$ and consider the points 
$$
p_{s,j}=\frac{j}{2^{s+1}}=\frac{j}{1+d(s)},\;\;\;j=1,...,d(s).
\eqno(10)
$$
For given total sample size $N=m+n$, we shall restrict attention to $s=S(N)$, where $S(N)$ is a user-defined increasing sequence, while $j=1,...,D(N)$, where $D(N)=d(S(N))$.

Also note that by changing $s=0,1,...$ in (10), and pertinent values of arguments of (8) and (9), we allow for more and more careful check of the discrepancy between $G$ and $F$.  More precisely, in the case of ${\sf U}_N$, we start with measuring the deviation at the empirical median of $F$, then we proceed to control a fit at the three empirical quartiles of $F$, etc. In the case of ${\sf P}_N$ similar interpretation applies with  quantiles of $H_N$ in place of the respective quantiles of $F_n$. In short, on each stage $s$ we are practically doubling the number of inspection points. Moreover, in contrast to many other papers where the grid is restricted to relatively small number of points, e.g. quintiles or deciles, we consider the size of the grid to be dependent on $N$. This allows to construct consistent tests. Obviously, other partition systems, if convenient, could be consider as well. 

Given $s$ and $d(s)$ in (10), evaluating ${\sf U}_N$ and ${\sf P}_N$ at points $p_{s,j},\; j=1,...,d(s),$  and representing resulting values  as bars over the grid points yields pertinent {\it bar plots} (B-plots), as labeled in Ducharme and Ledwina (2022), and introduced and exploited earlier, in the case of the process ${\sf P}_N$, in Ledwina and Wy{\l}upek (2012a,b). We refer to the two last mentioned papers for several examples of such B-plots for ${\sf P}_N$.

B-plots, as defined above,  are selected finite dimensional realizations of the processes ${\sf U}_N$ and ${\sf P}_N$. In the next section we shall show another interpretation of them, linking B-plots to score vectors in some auxiliary models for data at hand. 

The approach to testing in which the grid of points is taken into account has long history, going back to classical Pearson's test. It is also consistent with long lasting tradition in econometrics; cf. Anderson (1996) and  Davidson and Duclos (2000) for an illustration, and Whang (2019) for a detailed discussion. In econometrics, there is a certain tradition of considering relatively small number of inspection points. Such strategy  is good under some type of deviations from a null model but appears to be fatal under some other types. The paper by Lean et al. (2006) discusses this question and suggests some more careful inspection. See also Sriboonchitta et al. (2009) for related comments.  Alternative approach of comparing two groups via selected quantile differences have been proposed by Wilcox (1995) and applied latter on to analysis of medical data; see Rousselet et al. (2017) for an extensive overview. As mentioned above,  Ledwina and Wy{\l}upek (2012a,b), in the pooled case, have introduced diadic grid and related bars, and elaborated some of their applications.

When carefully designed,  approaches via comparison on a selected grid simplify interpretations and  are sufficient techniques to understand how the two groups of observations differ. For more comments see below.\\

\noindent
{\bf 4. Fourier coefficients of comparison densities  and their empirical versions}\\

\noindent
Our starting point to introduce CC and CCC were considerations on convenient modeling of a net of local alternatives in the two-sample problem. For example, Janic-Wr\'oblewska and Ledwina (2000), in the pooled setting,  have proposed such approach with use of exponential family pertinent to orthonormal Legendre polynomials. For related investigation in the unpooled case see Zhou et al. (2017). In the discussed pooled case, the resulting score vector consists  then of empirical Fourier coefficients of the function $c$, given in (4), in the polynomial system. There are two drawbacks of such modeling using polynomials. First, only few  first empirical Fourier coefficients have a meaningful interpretation helping to describe some differences in between the two populations. Second, polynomials are not handy when one likes to model one-sided alternatives. Therefore, in Ledwina and Wy{\l}upek (2012b) a new system of functions has been introduced.  These functions arise as normalized orthogonal projections of the Haar functions onto the cone of nondecreasing functions. Given the resolution level  $s$ and related dimension $d(s)$ the construction results in the $d(s)$-dimensional vector of functions $(h_{s,1}(p),...,h_{s,d(s)}(p))$, where
$$
h_{s,j}(p) = 
\frac{p_{s,j}-\mathbbm{1}\{0\leq p \leq p_{s,j}\}}{\sqrt{p_{s,j}(1-p_{s,j})}},\;\;j=1,...,d(s),
$$
while $\mathbbm{1}\{E\}$ denotes the indicator of the event $E$. It can be seen that 
$$
\int_0^1 r(p)h_{s,j}(p)\mu(dp)= \frac{p_{s,j}- G\bigl(F^{-1}(p_{s,j})\bigr)}{\sqrt{p_{s,j}(1-p_{s,j})}}
$$
and
$$ 
\int_0^1 c(p)h_{s,j}(p)\mu(dp)= \frac{F(H^{-1}(p_{s,j}))-G(H^{-1}(p_{s,j}))}{\sqrt{p_{s,j}(1-p_{s,j})}}.
$$
This shows that CC and CCC curves are aggregated theoretical Fourier coefficients, in the system $\{h_{s,j}\}$, pertinent to  $r(p)$ and $c(p)=r_1(p)-r_2(p)$, respectively. Similarly, $\widehat{CC}$ and $\widehat{CCC}$ can be viewed as the corresponding aggregated empirical Fourier coefficients, or scores, using standard terminology and standard way of building some pertinent local models. This means that B-plots can be interpreted as  graphical representations of pertinent score vectors. The last observation opens a way to construct, in the two-sample and analogous $k$-sample setting,  score-type tests with easily interpretable components, and thus extends a range of possible applications of B-plots. For example, Ledwina and Wy{\l}upek (2012a,b) and Wy{\l}upek (2013, 2016) have used the bars to construct data driven score-type tests. Data driven tests are, roughly speaking, quadratic forms of score vectors with the number of summands defined via a selection rule. Such construction solves the difficult question how many summands in the form to take. It allows to consider increasingly  finer grid, as $N$ increases, and yields the solution which adjust the number of summands to data at hand. Case 3 in Tables I-A and I-B of Barrett and Donald (2003) well illustrates how heavily the chosen size of the grid can influence pertinent empirical powers of quadratic statistics under the consideration there. This clearly supports importance of careful data driven constructions.

In the present paper we focus on alternative way of combining the scores, already considered in Ledwina and Wy{\l}upek (2012a). Namely, we consider some of their minima. This point is discussed below. For an illustration, we focus here on standard problem on stochastic dominance, while some other problems are briefly mentioned in Sections 5.5 and  Appendix E.\\

\noindent
{\bf 5. Testing for  stochastic dominance}\\

\noindent
Stochastically ordered families of distributions have been introduced in mathematical statistics  to study properties of power functions of some tests; cf. e.g. Lehmann (1955). However, since that time this order among distributions is being used in many areas of applications, such as reliability, econometrics, and clinical trials. This raises a need for good tools to validate such assumption.

Verification of
$$
{\mathbb H} : F(x) \geq G(x),\;\;\;\mbox{for all}\;\;\;x \in \mathbb R,
$$
against
$$
{\mathbb A} : F(x) < G(x),\;\;\;\mbox{for some}\;\;x \in \mathbb R,
$$
is the most popular formulation of testing problem in the context of detecting stochastic dominance. If ${\mathbb H}$ holds true we say that $G$ stochastically dominates over $F$. For an extensive review of this problem see Whang (2019), Section 2.2,  and Garc\'ia-G\'omez et al. (2019). \\

\newpage
\noindent
{\sf 5.1. Test statistics}\\[-1em]

For the aforementioned testing problem, under the pooled setting, Ledwina and Wy{\l}upek (2012a) have introduced and studied test statistic $M_{D(N)}$ which, up to the continuity correction, is equal to ${\cal P}_{D(N)}$, where
\vspace{-1em}
\begin{align*}
{\cal P}_{D(N)} &=\min_{1 \leq j \leq D(N)} {\sf P}_N(p_{S(N),j}) =\eta_N \times \min_{1 \leq j \leq D(N)} \widehat{\text{CCC}}(p_{S(N),j})\\
                &=\eta_N \times \min_{1 \leq j \leq D(N)}\frac{F_m(H_N^{-1}(p_{S(N),j}))-G_n(H_N^{-1}(p_{S(N),j}))}{\sqrt{p_{S(N),j}(1-p_{S(N),j})}}.
\end{align*}

The test  rejects ${\mathbb H}$ in favor of ${\mathbb A}$ if ${\cal P}_{D(N)}$ is too small. 
The statistic ${\cal P}_{D(N)}$ can be seen as a refined variant of the solution proposed by Anderson (1996). For related discussion see Ledwina and Wy{\l}upek (2012a), Section 3.
Ledwina and Wy{\l}upek (2012a) have shown that ${M}_{D(N)}$ is stochastically increasing under ${\mathbb H}$ and the error of the first kind attains maximum when $F = G$. Also, they have proved that if $D(N)=o(N)$ and $D(N) \to \infty$, as $N \to \infty$, than the test is consistent under ${\mathbb A}$. For practical implementation they recommended to apply the largest $D(N)$ do not exceeding $N$. Their extensive simulation experiments, reported in Ledwina and Wy{\l}upek (2012a, 2013), have shown that finite sample behavior of $M_{D(N)}$ is very stable under such choice of $D(N)$, even under considerably large $N$. The difference in the definitions of $M_{D(N)}$ and ${\cal P}_{D(N)}$ is cosmetic and ${\cal P}_{D(N)}$ inherits both finite sample as well as asymptotic properties after $M_{D(N)}$.

It is obvious that similar solution to ${\cal P}_{D(N)}$  can be proposed in the unpooled case. Such analog ${\cal U}_{D(N)}$, say,   takes the form 
\vspace{-1em}
\begin{align*}
{\cal U}_{D(N)} &=\min_{1 \leq j \leq D(N)} {\sf U}_N(p_{S(N),j}) = \eta_N \times \min_{1 \leq j \leq D(N)} \widehat{\text{CC}}(p_{S(N),j})\\
				&=\eta_N \times \min_{1 \leq j \leq D(N)}\frac{p_{S(N),j}-G_n(F_m^{-1}(p_{S(N),j}))}{\sqrt{p_{S(N),j}(1-p_{S(N),j})}}.
\end{align*}

Again, small values of $\;{\cal U}_{D(N)}$ are significant. If one is interested is studying  ROC(p) or ODC(p) over prescribed subinterval $(p_1,p_2)$ of (0,1), as is quite often the case in practice, the  statistic ${\cal U}_{D(N)}$  can be immediately adjusted to such situation.

In Appendix B we show that $\;{\cal U}_{D(N)}$, similarly as ${\cal P}_{D(N)}$, is stochastically increasing under ${\mathbb H}$ and the error of the first kind attains maximum when $F = G$. Under $F = G$ asymptotic distributions of selected group of  bars in both settings are the same, cf. Appendix A. However, finite sample results on $\;{\cal U}_{D(N)}$ and  ${\cal P}_{D(N)}$ under $F = G$ are not the same; cf. Section 5.4 and Appendix C. Moreover, from existing results, see Appendix A, it follows that when $F \neq G$ the process ${\sf U}_N$ is expected to have, in many important practical situations,  larger variance than the process ${\sf P}_N$. This, in particular,  influences finite sample behavior of $\;{\cal U}_{D(N)}$ under alternatives. Some instability of ${\sf U}_N$ near ends of (0,1) is in agreement with finite sample behavior of standard ROC estimate, which we use via the empirical ODC. For related comments on variability of empirical ROC curve see Gon\c{c}alves et al. (2014). Obviously, the standardization which we introduce in the process ${\sf U}_N$ additionally increases the inherited variability. On the other hand, this process clearly exhibits the real scale of the problem. In Appendix B, consistency of $\;{\cal U}_{D(N)}$ is proved when $D(N)=o(N/\log^3 N)$. Although, when it comes to $D(N)$, consistency results on $\;{\cal U}_{D(N)}$ and ${\cal P}_{D(N)}$ are not very different, our simulations show that one has to be careful with a choice of $D(N)$ in $\;{\cal U}_{D(N)}$, as empirical critical values of $\;{\cal U}_{D(N)}$ are much more sensitive to the choice of $D(N)$ than corresponding critical  values of $\;{\cal P}_{D(N)}$.

Taking into account these observations and having in mind statistical practice, we have decided to apply in the main body of this contribution, both in real data examples and in simulations, $D(N)=127$ in both statistics $\;{\cal U}_{D(N)}$ and ${\cal P}_{D(N)}$. Additionally, in real data analysis, this size of the grid is accompanied  with acceptance regions, see below,  pertinent to deciles of respective reference distributions. It seems that such strategy provides sufficiently careful inspection of possible deviations from the null model in a standard analysis. Additionally, such choice reduces  inherited variability pertinent to the unpooled approach.

To have more complete picture of the situation, in Appendix C we illustrate an influence of increasing $D(N)$ onto empirical  null distributions of pertinent test statistics and acceptance regions. \\

\noindent
{\sf 5.2. Acceptance regions for subsets of bars in B-plot}\\[-1em]

\noindent
As mentioned in Section 1, our purpose is not only to construct sensitive tests but also to provide an accompanying tool that can help understand what is decisive for the possible rejection or acceptance of the null hypothesis. For this purpose, following the development of Ducharme nad Ledwina (2022), we propose to consider acceptance regions for some selected subsets of bars.

In case of testing $\mathbb H$ against $\mathbb A$ small negative values of bars indicate a possible disagreement with $\mathbb H$. Due to (7), it is almost immediately seen if a single bar looks like to be significantly small. When thinking on significance of a group of bars, in view of their correlations,  we have to be careful. However, due to Corollary B.1  of Appendix B, it is easy to control the situation.  Therefore, we propose the following procedure.

\noindent
${\bullet}$ Consider ten intervals $I_1=[0,0.1],\;I_2=(0.1,0.2],...,I_{10}=(0.9,1]$. \\
${\bullet}$ Given $N$ and $D(N)$, find subsets of points $p_{S(N),j}$ which fall into succeeding intervals $I_1,...,I_{10}$.\\
${\bullet}$ Under $F = G$, for both processes ${\sf U}_N(p)=\eta_N \widehat{\text{CC}}(p)$ and ${\sf P}_N(p)=\eta_N \widehat{\text{CCC}}(p)$, $\;p \in (0,1)$, and  
a given interval $I_k$, say, calculate the barriers $l_{\sf U}(N,\alpha, I_k)$ and $l_{\sf P}(N,\alpha, I_k)$ defined as follows
$$
P\Bigl(\min_{j \in I_k} {\sf U}_N(p_{S(N),j}) \geq l_{\sf U}(N,\alpha,I_k)\Bigr) \geq 1-\alpha,
$$
and 
$$
P\Bigl(\min_{j \in I_k} {\sf P}_N(p_{S(N),j}) \geq l_{\sf P}(N,\alpha,I_k)\Bigr) \geq 1-\alpha.
$$ 
\\
We propose to represent these one-sided simultaneous $1-\alpha$ level acceptance regions by shaded strips on B-plots, see Section 5.4. For notational convenience we set
$$
L({\sf U}_N,I_k)=\min_{j \in I_k} {\sf U}_N(p_{S(N),j})\;\;\;\mbox{and}\;\;\;L({\sf P}_N,I_k)=\min_{j \in I_k} {\sf P}_N(p_{S(N),j})
\eqno(11)
$$
to denote the local minima of both processes. Throughout we consider $\alpha=0.05$. Since, under $F=G$, the processes ${\sf U}_N$ and ${\sf P}_N$ are distribution free, the barriers can be easily found by ordinary Monte Carlo simulations. Throughout the paper the number of MC runs is $100\;000$.

In other testing problems, similarly defined,  upper or two-sided $\alpha$-level acceptance regions can be of interest. Note also that acceptance regions of the proposed form can be applied to provide clear hint upon possible invalidity  of the null hypothesis, in spite of the form of overall test statistic used for its verification. 

	It should be emphasized that we propose and use the acceptance regions to evaluate  some local discrepancies between the two samples. As highlighted in Section~1, such question is of vital interest in recent years.\\

\noindent
{\sf 5.3. Simulation study}\\[-1em]

\noindent
In this section, we investigate the performance of the tests $\;{\cal U}_{D(N)}$ and ${\cal P}_{D(N)}$ using Monte Carlo simulations. In view of existing evidence on $M_{D(N)}$, a forerunner of ${\cal P}_{D(N)}$, main goal of this study is to contrast pooled and unpooled approaches, and to show to which extent weighted variant of Kolmogorov-Smirnov test outperforms unweighted one.\\

\pagebreak

\noindent
{\sf  5.3.1. Alternatives}\\[-1em]

\noindent
We consider standard scenarios already considered in some simulation studies. Similarly as in Ledwina and Wy{\l}upek (2012a), each alternative is described as follows: ${\mathbb A}$: {\it a description of $F$/a description of }$G$. The list of models is as follows:\\[-1em]

\noindent
$\bullet$ ${\mathbb A}_1$ : $L(1)/L(0.7)$, where $L(\theta)$ is the CDF $[\Phi(\cdot)]^{\theta},\;$ while $\Phi(\cdot)$ is the CDF of N(0,1) random variable. This is a particular case of Lehmann's model which is popular in ROC applications; cf. e.g. G\"{o}nen and Heller (2010);\\
$\bullet$ ${\mathbb A}_2$ : $P(1)/P(1.3)$, where $P(\theta)$ stands for the Pareto CDF. Such standard income model has been used by Schmidt and Terede (1996);\\
$\bullet$ ${\mathbb A}_3$ : $\Psi(0.6)/Uniform(-1,1)$, where $\Psi(\mu)$ is the local alternative defined by Fan (1996), see Example 5 therein;\\
$\bullet$ ${\mathbb A}_4$ : $[(0.4)N(0.4,1)+(0.6)\chi^2_1]/N(0.4,1)$, where $N(\mu,\sigma^2)$ stands for the normal distribution with parameters $\mu$ and $\sigma^2$, while $\chi^2_1$ denotes the chi-square distribution with one degree of freedom, cf. Ledwina and Wy{\l}upek (2012a);\\
$\bullet$ ${\mathbb A}_5$ : $LN(0.85,0.6)/[(0.9)LN(0.85,0.4)+(0.1)LN(0.4,0.9)]$, cf. Barrett and Donald (2003), who motivated the choice by some evidence in income analysis;\\
$\bullet$ ${\mathbb A}_6$ : $A(0)/A(1.3)$, where $A(\theta)$ is Andreson's kurtotic distribution generated as $Z\cdot |Z|^{\theta}$, while $Z$ is N(0,1) and $\theta \geq 0$, cf. Anderson (1994);\\
$\bullet$ ${\mathbb A}_7$ : $N(0,1)/N(0,2.25)$,  the normal scale model with $\sigma=1.5$, one of the simplest models applied in ROC studies.\\
$\bullet$ ${\mathbb A}_8$ : $Laplace(0,1)/Laplace(1,2.5)$, where $Laplace(\mu,\sigma)$ is the standard location-scale Laplace family, cf. Ledwina and Wy{\l}upek (2012a);\\
$\bullet$  ${\mathbb A}_9$ : $LN(0.85,0.6)/LN(1.2,0.2)$, next log-normal case considered by Barrett and Donald (2003), again with income analysis perspective.\\[-1em]

\begin{figure}[h!]
	\centering
	\includegraphics[width=.75\linewidth]{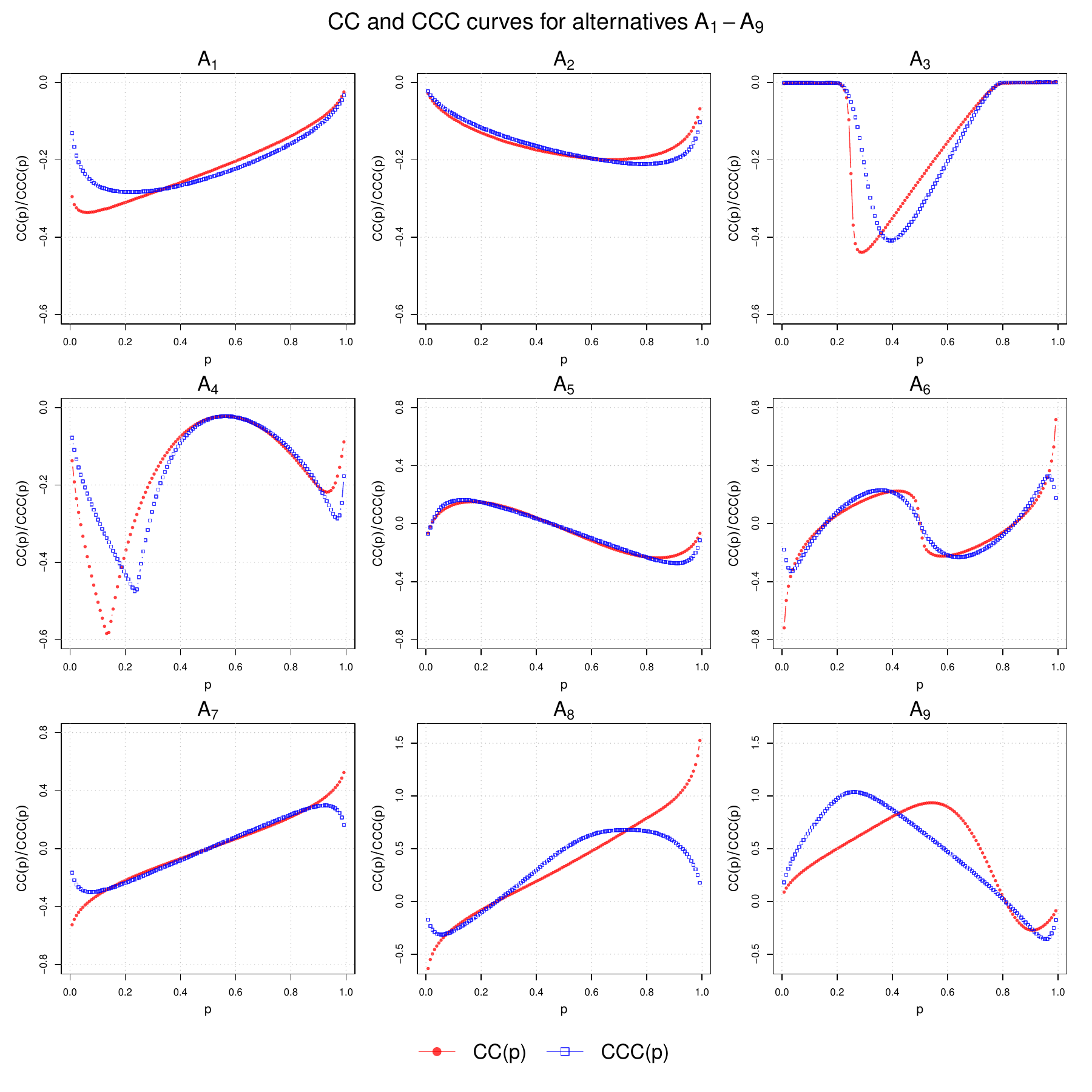}
	\caption{Graphical representation of alternatives ${\mathbb A}_1 - {\mathbb A}_9$. MC estimated CC -- red dots, and MC estimated CCC -- blue squares, over the grid of  127 points.}
	\label{Fig:1}
\end{figure}

The cases ${\mathbb A}_2, {\mathbb A}_4, {\mathbb A}_5, {\mathbb A}_8,$ and ${\mathbb A}_9$ 
have been already considered in the extensive simulation study in Ledwina and Wy{\l}upek (2012a). There the attention has been focused on  $M_{D(N)}$
and its relation to a several classical tests and new data driven test. Alternatives were graphically presented by respective sets of bars pertinent to $\text{CCC}(p)$ plots, over 32 partition points, using the notation of the present paper. For similar illustrations see Ledwina and Wy{\l}upek (2012b, 2013).

Here, in Figure 1,  we present the considered alternatives  by showing pertinent (MC estimated) CC and CCC curves. The MC estimates have resulted from averaging of the simulated results for $\widehat {\text{CC}}$ and $\widehat {\text{CCC}}$, in each of 127 points of the grid, under $m=n=5\; 000$ and  1 000 Monte Carlo runs. The subsequent panels of Figure 1 show that even such standard and relatively simple models yield large variety of shapes. Also, some differences between the two curves are sometimes striking.

Below, we study how the above-defined tests and two classical solutions react to such alternatives.\\

\noindent
{\sf 5.3.2. Competitors and empirical powers}\\[-1em]

\noindent
We compare $\;{\cal U}_{D(N)}$ and ${\cal P}_{D(N)}$ to two well-established standards in the field of stochastic dominance: adjusted variants of Kolmogorov-Smirnov test, denoted by ${\cal K}_N$, and AUC-type statistic, denoted by ${\cal T}_N$; for more details see e.g. Barrett and Donald (2003), and Schmid and Trede (1996). Both statistics are briefly recalled below.

As in Barrett and Donald (2003), we consider
$$
{\cal K}_N=\eta_N \times \sup_{x \in \mathbb R}  \{G_n(x)-F_m(x)\}.
$$
${\mathbb H}$ is rejected for large values of ${\cal K}_N$. The critical value $k(\alpha;N)$ is defined by $P({\cal K}_N \geq k(\alpha;N);F=G) \leq \alpha.$ For an event $E$, the notation $P(E;F=G)$ denotes the probability of $E$ calculated under $F=G$.

In turn, an AUC-type statistic defined by Schmidt and Trede (1996) has the form
$$
{\cal T}_N=\eta_N \times \int_{\mathbb R}\{G_n(x)-F_n(x)\}^{+}dF_n(x)=
\eta_N \times \Big\{ \frac{1}{m}\sum_{i=1}^m \Big[G_n(X_{(i)})-\frac{i}{m}\Big]^{+}\Big\},
$$
where $z^+=\max\{z,0\}$. Large values of ${\cal T}_N$ are significant and the corresponding critical region is given by $\{{\cal T}_N \geq t(\alpha;N);F=G\}$, where $t(\alpha;N)$ stands for the related critical value. 

Throughout the paper the significance level is always 0.05, while related  critical values are always based on $100\;000$ MC runs. For all alternatives listed above, $m=n=120$ and $D(N)=127$, we calculated empirical powers of the four tests on the basis of 5 000 MC runs.
Obtained results are reported in Table 1. Simulated critical values of $\;{\cal U}_{D(N)}$, ${\cal P}_{D(N)}$, ${\cal K}_N$ and ${\cal T}_N$ are as follows: -3.188, -2.764, 1.162, and 0.457, respectively. 

\begin{table}[h!]
\caption{Empirical powers of the four tests  using 5 000 Monte Carlo samples generated from models ${\mathbb A}_1 - {\mathbb A}_9$; $m=n=120, D(N)=127$.}
\label{Tab:1}
\begin{center}
\begin{tabular}{rrrrrrrrrr} \hline
Test & ${\mathbb A}_1$ & ${\mathbb A}_2$ &${\mathbb A}_3$  &${\mathbb A}_4$ &${\mathbb A}_5$ &${\mathbb A}_6$ &${\mathbb A}_7$ &${\mathbb A}_8$ & ${\mathbb A}_9$ \\
\hline
${\cal U}_{D(N)}$ &{\sf 74}	&27 &71 &{\sf 91} &11 &{\bf 84} &{\bf 79} &{\bf 80}	&4  \\
${\cal P}_{D(N)}$ &73 &47 &{\sf 79}	&{\bf 93}	&{\bf 53} &{\sf 70}	&{\sf 67} &{\sf 65} &{\bf 67} \\
${\cal K}_N$      &71  &{\sf 50} &{\bf 88} &87	&{\sf 35} &42 &31 &11 &{\sf 7}  \\
${\cal T}_N$      &{\bf 77}	&{\bf 53} &58 &53 &17 &10 &13 &0 &0  \\
\hline
\end{tabular}

\vspace{1em}
{\footnotesize  The { boldface} and { Sans Serif} values represent the highest and second highest powers.}
\end{center}
\end{table}
\vspace*{-.5\baselineskip} 

Table 1 shows that empirical powers of $\;{\cal U}_{D(N)}$ are not stable. Though they are very competitive in some cases, in other ones they are poor. ${\cal P}_{D(N)}$ has definitely more stable behavior and outperforms considerably the competitors in this regard. Also, values of empirical powers of ${\cal P}_{D(N)}$ are satisfactory. Both tests are based on minimal values of the respective estimated comparison curves. Evidently, greater stability of $\widehat{\text{CCC}}$ plays crucial role in ensuring more satisfactory empirical powers of pertinent test statistic ${\cal P}_{D(N)}$.   Our experiments indicate that aiming at stable construction of acceptance regions and tests on the basis of ${\sf U}_N$ requires presumably some smoothing of this process and more careful choice of pertinent test statistic. However, such investigations are out of the scope of the present contribution.

As to the powers of ${\cal K}_N$ and ${\cal T}_N$, the results in Table 1 are as expected. They are consistent with earlier evidence and existing theoretical results. In particular, see simulation results in Ledwina and Wy{\l}upek (2012a, 2013), efficiency considerations in Inglot et al. (2019), and references therein. It is worth of noticing that, neglecting any formal statements, we can formulate some conclusions on the expected empirical power of ${\cal K}_N$ and ${\cal T}_N$ simply by looking at the plot of CC curve. For example, the statistics ${\eta_N}^{-1} {\cal K}_N$ estimates $S=\sup_{x \in {\mathbb R}}\{G(x)-F(x)\}=\sup_{[0,1]} \{-\sqrt{p(1-p)}\text{CC}(p)\}$. Hence, if CC$(p)$ is negative and small near the center of [0,1] than $S$ is large and it is relatively easy to reject ${\mathbb H}$, under moderately large $m$ and $n$. Simply, then asymptotic shift is reasonably large. In contrast, when CC$(p)$ indicates visible discrepancies near 0 or 1, then the term $\sqrt {p(1-p)}$ considerably reduces this signal and rejection of ${\mathbb H}$, basing on samples of the same  sizes as in the previous situation, is much harder. Similarly, using corresponding integral, one can interpret probable response of ${\cal T}_N$ on a particular shape of CC.  Reasoning of this kind  provide also some  guidance
whether the classical tests ${\cal K}_N$ and ${\cal T}_N$ are likely to be sensitive enough in a given situation.\\

\noindent
{\sf 5.4. Real data examples}\\

\noindent
{\sf 5.4.1. Income data analysis}\\[-1em]

\noindent
In this section, we shall illustrate the usefulness of the approaches discussed in our paper in the context of real-world problem related to the income distribution comparison. Specifically, the dataset under consideration concerns after-tax incomes in Canada in 1978 and 1986, and comes from the Canadian Family Expenditure Survey. These data were previously analyzed by Barrett and Donald (2003), among others. 

For an illustration,  we present results related to 1978 versus 1986 comparison, using  Barrett and Donald (2003) terminology.
Thus,  we will investigate whether  the (unknown) CDF G in 1978 is less than or equal to the (unknown) CDF F, referring to 1986 income. 
In other words, we inquire if the after-tax income in 1986 is stochastically smaller than the pertinent income in 1978.
The data consists of $n = 8\; 526$ observations from 1978 and $m = 9\; 470$ observations from 1986. Hence, $N=17\; 996$. 

Our analysis begins by applying the proposed tests $\;{\cal U}_{D(N)}$ and $\;{\cal P}_{D(N)}$, and  comparing them with the classical competitors, i.e.\ $\;{\cal K}_{N}$ and $\;{\cal T}_{N}$. 
Here we calculate $\;{\cal U}_{D(N)}$ and $\;{\cal P}_{D(N)}$ taking $S(N)=6$ and $D(N)=127$. We have considered $\alpha=0.05$. Numerical results are summarized in Table 2. Related results for $S(N)=13$ and $D(N)=16\; 383$ are reported in Appendix C.1. 

\begin{table}[h!]
	\renewcommand\thetable{2}
	\caption{Basic outcomes in  Canadian after-tax family income in 1978 versus 1986 comparison with $m=9\;470,\;n=8\;526$ and $D(N)=127$.}
	\label{Tab:2}
	\begin{center}	
		\begin{tabular}{l|cccc}
			\hline
			Test statistic	& $\mathcal{U}_{D(N)}$ & $\mathcal{P}_{D(N)}$ & $\mathcal{K}_N$ &  $\mathcal{T}_N$\\
			\hline
			Obtained value  &-7.500 &-6.790 &1.888 &0.466  \\
			Critical value  &-2.854 &-2.848 &1.222 &0.482  \\
			p-value	       &0      &0      &0.001 &0.056\\
			\hline
		\end{tabular}
	\end{center}
\end{table}

\vspace*{-.5\baselineskip} 

The results in Table 2 indicate that, on the significance level 0.05, the assertion that the distribution function of incomes in 1978 is less or equal to the corresponding distribution in 1986 is evidently rejected by ${\cal U}_{D(N)}$, ${\cal P}_{D(N)}$, and ${\cal K}_N$, while  ${\cal T}_N$  accepts such hypothesis. To better understand the sources of these decisions, we will now employ appropriate graphical tools.

First, to have a closer look at the situation, we present rescaled difference in empirical CDF's (see top left plot in Figure~2), i.e. $\eta_N[G_n(\cdot)-F_m(\cdot)]$, where  $\eta_N=\sqrt{mn/N}$. 
Note that measuring such deviations is the basis for both considered classical tests, i.e. ${\cal K}_N$ and  ${\cal T}_N$. 
More precisely, $\;{\cal K}_{N}$ is the maximum of this curve, while ${\cal T}_N$ is the area under this curve and above 0 level.  
On the other hand,  ${\cal U}_{D(N)}$ and ${\cal P}_{D(N)}$  summarize the behavior of empirical processes ${\sf U}_N(p)$ and ${\sf P}_N(p)$ over the grid of 127 points, see top right plot in Figure~2, by showing their minima.

Let us recall that visual inspections of realizations of processes ${\sf U}_N(p)$ and ${\sf P}_N(p)$ over a selected grid of points  may provide  useful information on the local discrepancies between compared distributions. In particular, it is very common that these tools clearly manifest differences between G and F appearing in tails.
For example, in our case, both  ${\sf U}_N(p)$ and ${\sf P}_N(p)$ indicate that the differences in the upper tail of the distributions are relatively large and can be statistically significant, which was not apparent when we restrict ourselves to examining differences in empirical CDF's (cf. top left plot in Figure~2). 
It should be also highlighted that such initial visual findings can be formally confirmed with the aid of corresponding acceptance regions. 

In order to gain insight into the statistical significance of the observed differences in distributions we show bar plots  pertinent to $D(N)=127$ along with simultaneous 95\% one-sided acceptance regions (see middle row in Figure~2). Numerical values of related barriers in acceptance regions for this example are given in Appendix C.1. 
As mentioned in Section~5.2, small negative  values of bars indicate a possible disagreement with actually tested stochastic order.  In our case, obtained B-plots and  acceptance regions  clearly show statistically  significant disagreement in both tails of the two income distributions under comparison. More precisely, in the ranges of the first and last deciles, the income distribution in 1986 (F) is significantly stochastically larger than the distribution in 1978 (G), what is evidently inconsistent with investigated (hypothetical) stochastic dominance of 1978 over 1986 incomes. 
Also, in the ranges of seven and eight deciles, as well as eight and nine deciles, hypothesized relation $G\leq F$ is significantly violated. For instance, let us consider in more detail 10\% left and right tails of the incomes. Significantly negative bars in $(0,0.1]$ indicate thinner left tail of F than G.
This implies that the considered 10\% of lowest incomes in 1986 were significantly larger than in 1978. 
In turn, significantly negative bars in $(0.9,1)$ imply that 10\% right tail of after-tax incomes in 1986 was fatter than related tail in 1978. Hence, in the range of 10\% highest incomes, in 1986 significant  increase is observed.
The above supports and further explains the decision on lack of acceptance by some overall procedures of the considered stochastic order hypothesis on the  significance level $\alpha=0.05$.
Recall, however, that  despite such strong evidence on local discrepancies of distributions, the test ${\cal T}_N$ fails to reject null hypothesis.

In addition to aforementioned formal inference, B-plots and acceptance regions, although originally expressed as functions of quantiles $p$, can also be useful for extracting important information related more directly to the original data.
For example, the observed discrepancies between both distributions for the lower 10\% region are apparently larger than corresponding differences observed for the upper 30\%.  Thus, for families with the lowest income, i.e. less than about \$10 200 (estimated 1st decile, in both pooled and unpooled comparisons), the observed (relative) increase in annual after-tax income, when comparing 1978 to 1986, is more pronounced than for those with higher incomes, i.e. above about \$37 000  (estimated 7th decile in both comparisons).

Finally, corresponding box plots are presented (see bottom row in Figure~2) obtained for distributions of $L({\sf U}_N,I_k)$ and $L({\sf P}_N,I_k),\; k=1,...,10$ (see Section 5.2), under $F = G$. Thus, as above, we restrict attention to discrepancies pertinent to $\;p_{S(N),j}$'s $\;$ falling into the subsets $I_1,...,I_{10}$, defined by succeeding deciles of the corresponding reference distributions. 
Here, and throught the paper, box plots are based on 100 000 MC runs. 
In our case, for both unpooled and pooled comparisons  we obtained relatively stable results. However, this may not be the case for a different choice of $D(N)$ and/or for other datasets, where unpooled approach may exhibit significant instability (cf. corresponding results in Section 5.4.2 and in Appendix C).  \\

\begin{figure}[h!]
	\renewcommand\thefigure{2}
	\centering
	\includegraphics[width=.75\linewidth]{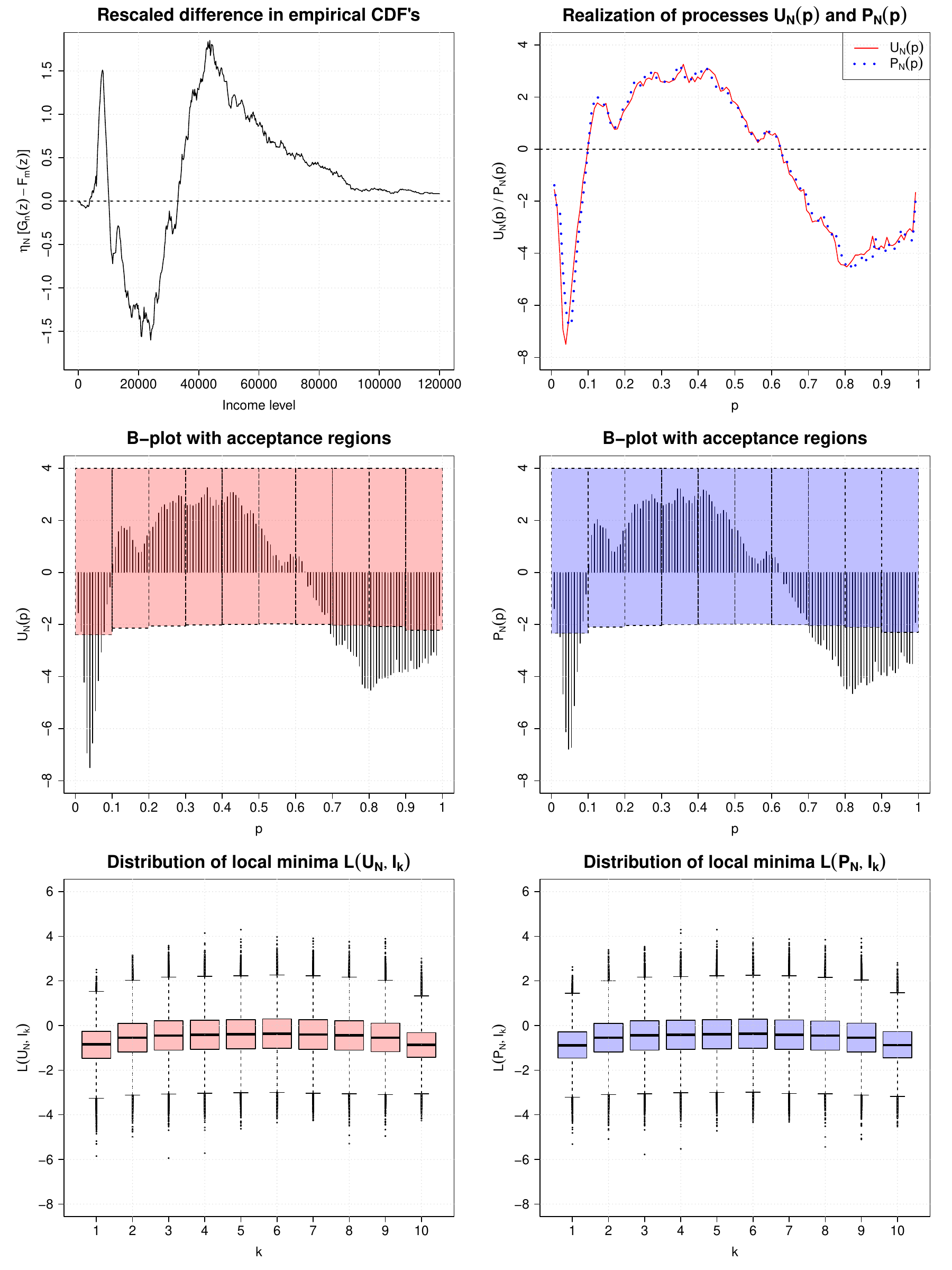}
	\caption{Results obtained  for  Canadian after-tax family income in 1978 versus 1986 comparison with $m=9\;470,\;n=8\;526$ and $D(N)=127$. The figure shows: rescaled difference in empirical CDF's (top-left plot), realizations of empirical processes ${\sf U}_N(p)$ and ${\sf P}_N(p)$  over the 127 points grid (top-right plot), B-plots along with 95\% one-sided simultaneous acceptance regions pertaining to  ${\sf U}_N(p)$ and ${\sf P}_N(p)$ (middle row),  and corresponding box plots (bottom row).}
	\label{Fig:2}
\end{figure}


\newpage

\noindent{\sf 5.4.2. Cholesterol data analysis}\\[-1em]

\noindent
Boon et al. (2013) have provided a three-dimensional visualization of empirical QQQ plots and the corresponding confidence tubes. They have applied their approach to cholesterol data collected by Diverse Population Collaboration Group. Also,  they  made available the data on the website accompanying their paper.
As one of the results of their study, Boon et al. (2013) have derived some  conclusions on stochastic ordering between the three population groups under their study. They investigated serum cholesterol level (in mg/dl) at baseline in men aged 45-65 years, living in Massachusetts, Honolulu, or Puerto Rico. 
Our approach to all possible two-sample problems in this study leads to similar or sometimes even more far-reaching conclusions on stochastic order as those formulated in Boon et al. (2013). 

For an illustration, we present results for subsample of obese men (BMI $> 30$) living in Honolulu and Puerto Rico. From one hand, it is known that obesity is associated with increased risk of several diseases. On the other, expected and observed by us differences in pairwise comparisons, for this stratum,  in the three populations, are least obvious. The selected comparison will be labelled as {\it Puerto Rico versus Honolulu}. Therefore, we will examine whether the (unknown) CDF G corresponding to cholesterol levels for obese men in Puerto Rico is less than or equal to the (unknown) CDF F relating to cholesterol levels for obese men living in Honolulu. Corresponding sample sizes are $m=160$, $\;n=628$, and hence $N=788$. Thus, in contrast to the previous example, we have much smaller  and relatively unbalanced  sample sizes. 

Similarly as in Section~5.4.1, we start the analysis by applying both  proposed tests  $\;{\cal U}_{D(N)}$ and $\;{\cal P}_{D(N)}$ as well as classical competitors, i.e.\ $\;{\cal K}_{N}$ and $\;{\cal T}_{N}$. 
We calculate $\;{\cal U}_{D(N)}$ and $\;{\cal P}_{D(N)}$ taking $S(N)=6$  and $D(N)=127$. Related results for $S(N)=8$ and $D(N)=511$ are reported in Appendix C.2. 
Numerical results for all the tests considered are summarized   in Table~3. 

\begin{table}[h!]
	\renewcommand\thetable{3}
	\caption{Basic outcomes in Puerto Rico versus Honolulu comparison of cholesterol levels in obese men groups with $m=160, n=628$ and $D(N)=127$. }
	\label{Tab:3}
	\begin{center}	
		\begin{tabular}{l|cccc}
			\hline
			Test statistic	& $\mathcal{U}_{D(N)}$ & $\mathcal{P}_{D(N)}$ & $\mathcal{K}_{N}$ &  $\mathcal{T}_{N}$\\
			\hline
			Obtained value	&-6.019 &-5.038 &1.244 &2.481 \\
			Critical value	&-3.491 &-2.813 &0.454 &1.205 \\
			p-value	    &0.003 &0 &0 &0 \\
			
			\hline
		\end{tabular}
	\end{center}
\end{table}

\vspace*{-.5\baselineskip} 

In this case, all four tests, on the chosen significance level 0.05, 
consistently reject null hypothesis that the cholesterol level distribution function of Puerto Rico's obese male population (G) is less than or equal to the corresponding cholesterol level distribution in Honolulu (F).
Going along the same lines as in Section~5.4.1, let us investigate the sources of departures from the null model with the aid of proposed graphical tools.

First, in Figure~3 we present rescaled difference in empirical CDF's (see top left plot) as well as  the realizations of empirical processes ${\sf U}_N(p)$ and ${\sf P}_N(p)$ over the grid of 127 points (see top right plot). We observe that these data are essentially differently distributed from the income data considered in Section~5.4.1. Moreover, it is seen that making a decision should be easier than in the previous situation, depicted in Figure~2, as both plots in the first row of Figure~3 suggest stronger violation of stochastic order relation. Note also that, in the present case, significant difference in the behavior of empirical processes   ${\sf U}_N(p)$ and ${\sf P}_N(p)$  is clearly manifested for $p \in (0,0.3)$, which may affect the inference based on unpooled and pooled settings, respectively.  On the other hand,  differences observed in the upper tail of cholesterol level distribution are relatively small and may not be statistically significant. 

In order to formally confirm our initial visual findings, let us look at  B-plots pertinent to $D(N)=127$  and processes  ${\sf U}_N(p)$ and ${\sf P}_N(p)$, as well as corresponding 95\% one-sided acceptance regions (see middle row in Figure~3). This time, in contrast to the income dataset,  the largest distortions between both  distributions are observed near the middle range of cholesterol level, where hypothesized relation $G\leq F$ is significantly violated. On the other hand,  in the ranges of the last two deciles, corresponding values of bars are above barriers, hence differences are not statistically significant. 
It is also worth mentioning that when considering reverse comparison, i.e. Honolulu versus Puerto Rico, the conclusions regarding cholesterol level distributions are even more pronounced. Specifically, we obtain that over the wide range of the first nine deciles  obtained differences are statistically significant, i.e. values of corresponding bars exceed barriers pertinent to upper acceptance regions  (corresponding results are available from authors upon request).
 This is more precise information than the one given in Boon et al. (2013), p. 77, for the obese strata. They conclude only that the difference between the distributions occur at lower cholesterol levels.

Summarizing the above, our approach 
leads to similar conclusions on stochastic order as those formulated in Boon et al. (2013). 
However, we believe that the sources of these decisions as well as a character of underlying departures from the null model are much easier to see, understand, and evaluate applying the graphical and inferential tools proposed in our paper. 

This example provides also successive evidence, following that in Table 1, that pooled approach yields more stable results than unpooled one. Observe that, in  the range of the first decile, the barriers corresponding to unpooled and pooled comparisons  differ markedly.  Numerical values of pertaining barriers in acceptance regions can be found in Appendix C.2. 
To further explore the differences, we show box plots  (see bottom row in Figure~3) obtained for distributions of the barriers  $L({\sf U}_N,I_k)$ and $L({\sf P}_N,I_k),\; k=1,...,10$, under $F = G$, and pertinent to the above displayed B-plots.
Although $D(N)=127$ only, the box plot for ${\sf U}_{N}$  exhibits considerable larger variability than for ${\sf P}_{N}$. This is in contrast to the previously considered example and may be, at least in part, due to relatively unbalanced sample sizes. 
Note, however, that for nearly balanced partitions, ${\sf U}_N$ can be unstable as well, when $D(N)$ is very large; cf. Appendix C.1.  
In general, further deterioration in stability of unpooled approach can be expected  if $D(N)$ is increased and this is indeed the case; see Appendix C.2 for 
additional illustration. 
In contrast to the above findings for the unpooled case, under  the same conditions, analogous objects in pooled setting are very stable.

\begin{figure}[h!]
	\centering
	\includegraphics[width=.75\linewidth]{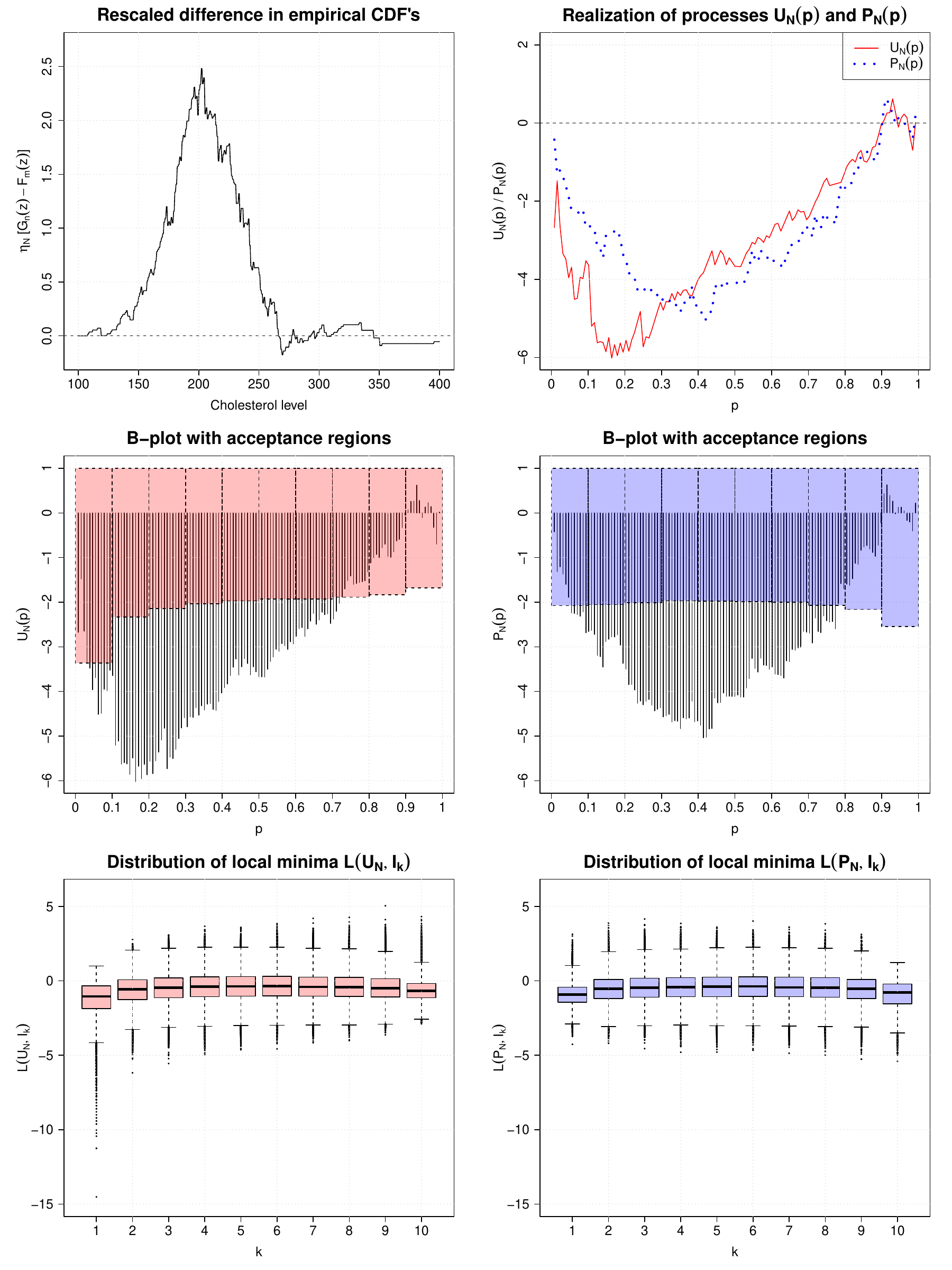}
	\caption{Results obtained  for comparison of cholesterol levels in obese men groups in Puerto Rico versus Honolulu with $m=160, n=628$ and $D(N)=127$.  The figure shows: rescaled difference in empirical CDF's (top-left plot), realizations of empirical processes ${\sf U}_N(p)$ and ${\sf P}_N(p)$  over the 127 points grid (top-right plot), B-plots along with 95\% one-sided simultaneous acceptance regions pertaining to  ${\sf U}_N(p)$ and ${\sf P}_N(p)$ (middle row),  and corresponding box plots (bottom row).}
	\label{Fig:3}
\end{figure}

\newpage

\noindent
{\sf 5.5. Other testing problems on stochastic dominance or some related trends}\\[-1em]

\noindent
In our paper, we have focused  on brief presentation of an application of comparison curves to one standard testing problem ${\mathbb H}$ against ${\mathbb A}$. However, similar approach can be applied to many other problems on stochastic dominance or some related trends. We mention some of the problems.

Davidov and Herman (2012) have considered testing ${\mathbb H}^0: F = G$ against stochastic dominance. Their solution exploits ODC and area under this curve. 

Testing the lack of stochastic dominance against stochastic dominance is a highly non-standard question. Davidson and Duclos (2013) have proposed two tests for so called restricted dominance. \'Alvarez-Esteban et al. (2017) considered approximate stochastic order to attack the problem. Both papers exploit bootstrap technique and provide tests on asymptotic level $\alpha$, being consistent against  sets of alternatives pertinent  to the imposed restrictions. However, it turns out that starting with ${\sf U}_{N}(p)$ can be useful in this and related problems. In particular, Ledwina and Wy{\l}upek (2012b) have proposed for testing the lack of dominance a data driven test on asymptotic level $\alpha$, being consistent under strict dominance, i.e. when $F > G$. The result has been sharpened in Wy{\l}upek (2013) to consistency under strict dominance and asymptotic unbiasedness otherwise. The paper also generalizes the solution to the $k$-sample problem. In Wy{\l}upek (2016) these results have been extended to detect an umbrella pattern.  Recently, Laha et al. (2022) have considered shape constrains in the case of testing the null of non-dominance against strict dominance.  For extensive additional information see Whang (2019).\\


\noindent
{\bf 6. Conclusions}\\

\noindent
We have discussed some relative distributions and proposed pertinent comparison curves which appear to be useful tools for investigating similarities and differences between two distributions. We have faced two settings: unpooled and pooled ones. In many applications areas, unpooled  samples and ODC/ROC are commonly used framework. In contrast, in mathematical statistics pooled samples and pertinent relative distributions are more popular. In both settings, we have introduced related comparison curves, denoted by CC and CCC, respectively,  and pertinent weighted empirical processes ${\sf U}_N$ and ${\sf P}_N$ on (0,1). All new objects have simple and intuitive interpretations. Hence, they are  potentially attractive for practitioners. 

We have focused on inference based on values of the processes on the grid of points. Representing obtained values as bars results in the B-plots.  B-plots, along with formally defined pertinent acceptance regions, allow to identify regions of the threshold $p \in (0,1)$ in which there are observed significant discrepancies between two samples. Also, B-plots can be interpreted in terms of scores and smooth components, or, equivalently, in terms of empirical Fourier coefficients of the comparison densities in the system of nondecreasing functions. This implies possible further applications of them. 

Our general finding from facing and contrasting the two parallel streams of facts and results pertinent to CC and CCC constructions and their standard empirical counterparts, in inter-distributional comparisons,  is as follows: the  pooled approach, pertinent to CCC curve, is much more convenient and reliable   than unpooled one, pertinent to CC and classical ROC/ODC estimates.

We have focused on one standard testing problem and on one method of combining the scores, as in the considered problem, min-type  statistics in the pooled setting were known to work very well. However, in many other problems data driven smooth tests or some other functions, pertinent to the bars, may be more adequate. The paper  Ducharme and Ledwina (2022) well exemplifies the duality.

Obviously, a possible application of relative distributions and pertinent weighted empirical processes is not restricted to different forms of stochastic dominance. For some evidence see Handcock and Morris (1999), p.~30, who list several settings in which relative distributions could be also used as a diagnostic tool.\\

\noindent
{\bf Acknowledgements}\\[-1em]

\noindent
We are grateful to Tadeusz Inglot for constructive remarks and suggestions. We would also like to thank Garry F. Barrett for providing the income data.\\


\noindent
{\bf References}\\ 
\setlist{nolistsep} 
\begin{spacing}{1}  
\begin{description}[topsep=0pt,itemsep=0pt,parsep=0pt,labelsep=0em]	
\item Algeri, S. (2021). Informative goodness-of-fit for multivariate distributions. {\it Electronic Journal of Statistics} {\bf 15}, 5570-5597.
\item Aly, E.-E.A.A., Cs\"{o}rg\H{o}, M., and Horv\'ath, L. (1987). P-P plots, rank processes, and Chernoff-Savage theorems. In: {\it New Perspectives in Theoretical and Applied Statistics}, M.L. Puri, J.P. Vilaplana, and W. Wertz, eds., Wiley, New York, pp. 135-156.
\item \'Alvarez-Esteban, P.C., del Barrio, E., Cuesta-Albertos, J.A., and Matr\'an (2017). Models for the assessment of treatment improvement: The ideal and the feasible. {\it Statistical Science} {\bf 32}, 469-485.
\item Anderson, G. (1994). Simple tests of distributional form. {\it Journal of Econometrics} {\bf 62}, 265-276.
\item Anderson, G. (1996). Nonparametric tests of stochastic dominance in income distributions. {\it Econometrica} 64, 1183-1193.
\item Arza, E., Ceberio, J., Irurozki, E., and P\'erez, A. (2023). Comparing two samples through stochastic dominance: A graphical approach. {\it Journal of Computational and Graphical Statistics} {\bf 32}, 551-566.
\item Bamber, D. (1975). The area above the ordinal dominance graph and the area below the receiver operating characteristic graph. {\it Journal of Mathematical Psychology} {\bf 12}, 387-415.
\item Barrett, G.F., and Donald, S.G. (2003). Consistent tests for stochastic dominance. {\it Econometrica} {\bf 71}, 71-104.
\item Beach, Ch.M., and Davidson, R. (1983). Distribution-free statsistical inference with Lorenz curves and income shares. {\it Review of Economic Studies} {\bf L}, 723-735.
\item Behnen, K. (1981). Nichtparametrische Statistik: Zweistichproben Rangtests. {\it Zeitschrift f\"{u}r Angewandte Mathematik und Mechanik} {\bf 61}, T203-T212.
\item Behnen, K., and Neuhaus, G. (1983). Galton's test as a linear rank test with estimated scores and its local asymptotic efficiency. {\it Annals of Statistics} 11, 588-599.
\item Behnen, K., and Neuhaus, G. (1989). {\it Rank Tests with Estimated Scores and Their Application}, Teubner, Stuttgart. 
\item Boon, M.A.A., Einmahl, J.H.J., and  McKeague, I.W. (2013) Visualizing multiple quantile plots. {\it Journal of Computational and Graphical Statistics} {\bf 22}, 69-78.
\item Carolan, Ch.A., and Tebbs, J.A. (2005). Nonparametric tests for and against likelihood ratio ordering in the two-sample problem. {\it  Biometrika} {\bf 92}, 159-171.
\item Cazals, F., and Lh\'eritier, A. (2015). Beyond two-sample-tests: localizing data discrepancies in high-dimensional spaces. {\it Research Report $n^o$ 8734}, Inria.
\item \'Cmiel, B., and Ledwina, T. (2020). Validation of association. {\it Insurance: Mathematics and Economics} {\bf 91}, 55-67.
\item Davidov, O., and Herman, A. (2012). Ordinal dominance curve based inference for stochastically ordered distributions. {\it Journal of the Royal Statistical Society, Ser. B.} {\bf 74}, 825-847.
\item Davidson, R., and Duclos, J.-Y. (2000). Statistical inference for stochastic dominance and for the measurement of poverty and inequality. {\it Econometrica} {\bf 68}, 1435-1464.
\item Davidson, R., and Duclos, J.-Y. (2013). Testing for restricted stochastic dominance. {\it Econometric Reviews} {\bf 32}, 84-125.
\item Doksum, K.A. (1974). Empirical probability plots and statistical inference for nonlinear models in the two-sample case. {\it Annals of Statistics} {\bf 2}, 267-277.
\item Doksum, K.A., and Sievers, G.L. (1976). Plotting with confidence: Graphical comparisons of two populations. {\it Biometrika} {\bf 63}, 421-434.
\item Ducharme, G., and Ledwina, T. (2022). A new set of tools for goodness-of-fit validation.\\ {\it arXiv:2209.07295v1 [stat.ME]}.
\item Duong, T. (2013). Local significant differences from nonparametric two-sample tests. {\it Journal of Nonparametric Statistics} {\bf 25}, 635-645.
\item Fan, J. (1996). Test of significance based on wavelet thresholding and Neyman's truncation. {\it Journal of the American Statistical Association} {\bf 91}, 674-688.
\item Fanjul-Hevia, A., and Gonz\'alez-Manteiga, W. (2018). A comparison study of methods for testing the equality of two or more ROC curves. {\it Computational Statistics} {\bf 33}, 357-377.
\item Garc\'ia-G\'omez, C., P\'erez, A., and Prieto-Alaiz, M. (2019). A review of stochastic dominance methods for poverty analysis. {\it Journal of Economic Surveys} {\bf 33}, 1437-1462.
\item Goldman, M., and Kaplan, D.M. (2018). Comparing distributions by multiple testing across quantiles or CDF values. Journal of Econometrics {\bf 206}, 143-166.
\item Gon\c{c}alves, L., Subtil, A., Oliveira, M.R., and de Zea Bermudes, P. (2014). ROC curve estimation: An overview. {\it REVSTAT - Statistical Journal} {\bf 12}, 1-20.
\item G\"{o}nen, M., and Heller, G. (2010). Lehmann family of ROC curves. {\it Medical Decision Making} {\bf 30}, 509-517.
\item Green, D.M., and Swets, J.A. (1966). {\it Signal Detection Theory and Psychophysics}, Wiley$\&$Sons, New York.
\item Handcock, M.S., and Morris, M. (1999). {\it Relative Distribution Methods in the Social Sciences}, Springer, New York.
\item Holmgren, E.B. (1975). The P-P plot as a method for comparing treatment effects. {\it Journal of the American Statistical Association} {\bf 90}, 360-365.
\item Huang, Y., and  Pepe, M.S. (2009). Biomarker evaluation and comparison using the controls as a reference population. {\it Biostatistics} {\bf 10}, 228-244.
\item Inglot, T., Ledwina, T., and \'Cmiel, B. (2019). Intermediate efficiency in nonparametric testing problems with an application to some weighted statistics, {\it ESAIM: Probability and Statistics} {\bf 23}, 697-738.
\item Janic-Wr\'oblewska, A., and Ledwina, T. (2000). Data driven rank tests for two-sample problem. {\it Scandinavian Journal of Statistics} {\bf 27}, 281-297.
\item Kiefer, J. (1959). K-sample analogues of the Kolmogorov-Smirnov and Cram\'er-v. Mises tests. {\it Annals of Mathematical Statistics} {\bf 30}, 420-447.
\item Kim, I., Lee, A.B., and Lei, J. (2019). Global and local two-sample tests via regression. {\it Electronic Journal of Statistics} 13, 5253-5305.
\item Krzanowski, W.J., Hand, D.J. (2009). {\it ROC Curves for Continuous Data}, Chapman and Hall, London.
\item Laha, N., Moodie, Z., Huang, Y., and Luedtke, A. (2022). Improved inference for vaccine-induced immune responses via shape-constrained methods. {\it Electronic Journal of Statistics} {\bf 16}, 5852-5933.
\item Lean, H.H., Wong, W.K., and Zhang, X.B. (2006). Size and power of some stochastic dominance tests: A Monte Carlo study. {\it SSRN Working Paper} No. 880988.
\item Ledwina, T. (2015). Visualizing association structure in bivariate copulas using new dependence function. In: {\it Stochastic Models, Statistics and Their Applications}, Springer Proceedings in Mathematics $\&$   Statistics 122, A. Steland et al. (eds),  pp. 19 - 27.
\item Ledwina, T., and Wy{\l}upek, G. (2012a). Nonparametric tests for first order stochastic dominance. {\it Test} {\bf 21}, 730-756.
\item Ledwina, T., and Wy{\l}upek, G. (2012b). Two-sample test for one-sided alternative. {\it Scandinavian Journal of Statistics} {\bf 39}, 358-381.
\item Ledwina, T., and Wy{\l}upek, G. (2013). Tests for first-order stochastic dominance. {\it Preprint IM PAN} 746.
\item Lehmann, E.L. (1955). Ordered families of distributions. {\it Annals of Mathematical Statistics} {\bf 26}, 399-419.
\item Li, G., Tiwari, R.C., and Wells, M.T. (1996). Quantile comparison functions in two-sample problem, with application to diagnostic markers. {\it Journal of the American Statistical Association} {\bf 91}, 689-698.
\item Ma, L., and Mao, J. (2019). Fisher exact scanning for dependency. {\it Journal of the American Statistical Association} {\bf 114}, 245-258.
\item Mathew, A. (2023). Quantile cumulative distribution function and its applications. {\it Communications in Statistics - Theory and Methods}, https://doi.org/10.1080/03610926.2023.2176716.
\item Nakas, C.T., Bantis, L.E., and Gatsonis, C.A. (2023). {\it ROC Analysis for Classification and Prediction in Practice}. CRC Press.
\item Neuhaus, G. (1987). Local asymptotics for linear rank statistics with estimated score functions. {\it Annals of Statistics} {\bf 15}, 491-512.
\item Parzen, E. (1977). Nonparametric statistical data sciences: A unified approach based on density estimation and testing for `white noise'. {\it Technical Report 47}, Statistical Science Division, State University of New York at Buffalo.
\item Parzen, E. (1998). Statistical methods mining, two sample data analysis, comparison distributions, and quantile limit theorems. In: {\it Asymptotic Methods in Probability and Statistics: A Volume in Honour of Mikl\'os Cs\"{o}rg\H{o}} (B. Szyszkowicz, ed.), North Holland, Amsterdam, pp. 611-617.
\item Pepe, M. (2003). {\it Statistical Evaluation of Medical Tests for Classification and Prediction}, Oxford University Press, Oxford.
\item Pyke, R., and Shorack, G.R. (1968). Weak convergence of a two-sample empirical process and a new approach to Chernoff-Savage theorems. {\it Annals of Mathematical Statistics} {\bf 39}, 755-771.
\item Rousselet, G., Pernet, C.R., and Wilcox, R.R. (2017). Beyond differences in means: robust graphical methods to compare two groups in neuroscience. {\it bioRxiv preprint} doi:\\ https://doi.org/10.1101/121079.
\item Schmid, F., and Trede, M. (1996). Testing for first-order stochastic dominance: a new distribution-free test. {\it Journal of the Royal Statistical Society. Series D (The Statistician)} {\bf 45}, 371-380.
\item Sriboonchitta, S., Wong, W.K., Dhompongsa, S., and Nguen, H.T. (2009). {\it Stochastic Dominance and Applications to Finance, Risk and Economics}, 
CRC Press, Boca Raton.
\item Thas, O. (2010). {\it Comparing Distributions}, Springer, New York.
\item Venkatraman, E. (2000). A permutation test to compare receiver operating characteristic curves. {\it Biometrics} {\bf 56}, 1134-1138.
\item Wang, D., Tang, Ch.-F., and Tebbs, J.M. (2020). More powerful goodness-of-fit tests for uniform stochastic ordering. {\it Computational Statistics and Data Analysis} {\bf 144}, 106898.
\item Whang, Y.-J. (2019). {\it Econometric Analysis of Stochastic Dominance}, Cambridge University Press, Cambridge.
\item Wieand, S., Gail, M.H., James, B.R., and James, K.L. (1989). A family of nonparametric statistics for comparing diagnostic markers with paired and unpaired data. {\it Biometrika} {\bf 76}, 585-592.
\item Wilcox, R.R. (1995). Comparing two independent groups via multiple quantiles. {\it Journal of the Royal Statistical Society. Series D (The Statistician)} {\bf 44}, 91-99.
\item Wy{\l}upek, G. (2010). Data-driven $k$-sample tests. {\it Technometrics} {\bf 52}, 107-123.
\item Wy{\l}upek, G. (2013). Data-driven tests for trend. {\it Communications in Statistics-Theory and Methods} {\bf 42}, 1406-1427.
\item Wy{\l}upek, G. (2016). An automatic test for the umbrella alternatives. {\it Scandinavian Journal of Statistics} 43, 1103-1123.
\item Xiang, S., Zhang, W., Liu, S., Hoadley, K.A., Perou, C.M.,  Zhang, K., and Marron, J.S. (2023). Pairwise nonlinear dependence
analysis of genomic data. {\it Annals of Applied Statistics} {\bf 17}, 2924-2943.
\item Zhang, W., Tang, L.L., Li, Q., Liu, A., and Lee, M.-L.T. (2018). Order-restricted inference for clustered ROC data with application to fingerprint matching accuracy. {\it Biometrics} {\bf 76}, 863-873.
\item Zhang, D.D., Zhou, X.-H., Freeman Jr.,  D.H., and Freeman, J.L. (2002). A non-parametric method for the comparison of partial areas under ROC curves and its application to large health care data sets. {\it Statistics in Medicine} {\bf 21}, 701-715.
\item Zhou, X.H., McClish, D.K., and Obuchowski, N.A. (2011). {\it Statistical Methods in Diagnostic Medicine}, Wiley$\&$Sons, New York.
\item Zhou, W.-X., Zheng, Ch., and Zhang, Z. (2017). Two-sample smooth tests for the equality of distributions. {\it Bernoulli} {\bf 23}, 951-989.
\end{description}
\end{spacing}
\newpage

\begin{center}
	{\Large{\bf Appendices}}
\end{center}


\vspace{0.2cm}


\vspace{0.2cm}


\noindent
{\bf Appendix A. On asymptotic distribution of bars}\\

\noindent 
In this section, we collect and discuss some existing results that help to understand joint distribution of bars in the two considered settings. Let us define two auxiliary (unweighted) processes
$$
{\sf A}_N(p)=\eta_N\Bigl\{G(F^{-1}(p))-G_n(F_m^{-1}(p))\Bigr\},\;\;\;p \in [0,1],
$$
and 
$$
{\sf B}_N(p)=\eta_N\Bigl\{F_m(H_{N}^{-1}(p))-G_n(H_{N}^{-1}(p)) - [F(H^{-1}(p))-G(H^{-1}(p))]\Bigr\},\;\;\;p \in [0,1],
$$
where $H_N(x)=\lambda_N F_m(x) + (1-\lambda_N) G_n(x),\;H(x)=\lambda_N F(x) + (1-\lambda_N) G(x)$, $\lambda_N=m/N$ and $\eta_N=\sqrt{mn/N}$. Also, as in the main body of the paper, $F_m$ and $G_n$ are empirical CDF's in the $X$ and $Y$ samples, respectively. 

In the case when $F = G$ the processes coincide  with the numerators of ${\sf U}_N$  and ${\sf P}_N$, respectively. 

The process ${\sf A}_N$ has been studied by Aly et al. (1987)  and Beirlant and Deheuvels (1990), among others.   Pyke and Shorack (1968) have investigated  ${\sf B}_N$.
We shall rely on the developments in Aly et al. (1987) and Pyke and Shorack (1968). Below, we formulate and comment on  some simple corollaries following from their results. Since we are concentrating on finite dimensional distributions of the processes (reweighted by $\sqrt{p(1-p)}$) one could wonder if an immediate derivation of such results would not be a better solution. It can be seen that finite dimensional distributions are related to linear rank statistics with non-continuous score generating functions. Such statistics are non-standard and some extra technical work is needed to elaborate the results, while our basic aim is to indicate existing differences between asymptotic distributions of vectors pertinent to the two processes. Working on technicalities is not central to this contribution.  Also, it is instructive to have wider perspective following from stochastic processes theory. 

To state these conclusions in a concise way, set 
$$
R(p)=G(F^{-1}(p)),\;\;\;R_1(p)=F(H^{-1}(p)),\;\;\;R_2(p)=G(H^{-1}(p)).
$$
Additionally, given $\lambda \in (0,1)$, introduce an auxiliary CDF $H_{\lambda}(x)=\lambda F(x)+(1-\lambda)G(x)$. Note also that $\eta_N={\sqrt N} {\sqrt{\lambda_N(1-\lambda_N)}}$. Recall  that $0 < \lambda_* \leq \lambda_N \leq 1-\lambda_* < 1, \;\lambda_* \leq 1/2$, and  that throughout we assume that $F$ and $G$ are continuous CDF's  and no ties are present among $X_1,...,X_m$ and $Y_1,...,Y_n$. Recall also that $r(p)=g(F^{-1}(p))/f(F^{-1}(p)),\;r_1(p)=dF(H^{-1}(p))/dp,\;r_2(p)=dG(H^{-1}(p))/dp,\; p \in (0,1)$.\\

\noindent
{\bf Proposition A.} 
\begin{enumerate}
	\item {\it Assume that $f$ and $g$ are continuous positive density functions on the open support of the distribution functions $F$ and $G$, respectively, and with some Chibisov-O'Reilly function $\omega$ we have $\sup_{0 \leq p \leq 1} [r(p) \omega (p)] < \infty$. Then, for any $p \in (0,1)$, it holds that}
	$$
	\eta_N\Bigl\{G_n(F_m^{-1}(p))-R(p)\Bigr\} 
	\sim AN\Bigl(0,\;\lambda_N R(p)[1-R(p)] + (1-\lambda_N) p(1-p)[r(p)]^2\Bigr),
	\eqno(A1)
	$$
	{\it where $AN(\cdot,\cdot)$ stands for: asymptotically normal with the indicated parameters. The second parameter is the asymptotic variance.}
	\item {\it Suppose that the functions $F(H_{\lambda}^{-1}(p))$ have derivatives $r_1(p;\lambda)=\partial F(H_{\lambda}^{-1}(p))/dp $, for each $p \in (0,1)$. Besides, for some ${\lambda}' \in (0,1), \; r_1(p;{\lambda}')$ is continuous on (0,1) and has one-sided limits at 0 and 1. Then, for any $p \in (0,1)$,}
	$$
	\eta_N\Bigl\{[F_m(H_{N}^{-1}(p))-G_n(H_{N}^{-1}(p))]-[F(H^{-1}(p))-G(H^{-1}(p))] \Bigr\}  
	$$
	$$
	\sim AN\Bigl(0,\; \lambda_N [r_1(p)]^2  R_2(p)[1-R_2(p)] + (1-\lambda_N)[r_2(p)]^2 R_1(p)[1-R_1(p)]\Bigr).
	\eqno(A2)
	$$
	
\end{enumerate}

\noindent
Similar conclusions can be formulated for finite dimensional distributions of the considered processes. Therefore, for completeness we also give here covariance functions of the processes ${\sf A}_N$ and ${\sf B}_N$, respectively. For any $p,q \in (0,1)$, it holds that
$$
E{\sf A}_N(p){\sf A}_N(q)=\lambda_N\Bigl[R(p) \wedge R(q) -R(p)R(q)\Bigr] + (1-\lambda_N)\Bigl[p \wedge q -pq\Bigr][r(p)r(q)],
\eqno(A3)
$$
where $x \wedge y$ stands for $\min\{x,y\}$; cf. p.~140 in Aly et al. (1987). In turn,
$$
E{\sf B}_N(p){\sf B}_N(q)=\lambda_N [r_1(p)r_1(q)]\Bigl[R_2(p) \wedge R_2(q)-R_2(p)R_2(q)\Bigr] + 
$$
$$
(1-\lambda_N)[r_2(p)r_2(q)]\Bigl[R_1(p) \wedge R_1(q)-R_1(p)R_1(q)\Bigr].
\eqno(A4)
$$
(A4) is a consequence of (3.9) in Pyke and Shorack (1968). However, note that there is a misprint in this formula and it should be multiplied by $\sqrt {1-\lambda_N}$; cf. (17) in Pyke (1970), and see also the comment below (1.15) in Aly et al. (1987) and their Theorem 3.3.

It is worth of noticing that both processes have nonnegative asymptotic covariance functions. 

The assumption in the point 2. of Proposition A coincides with Assumption 4.1 in Pyke and Shorack (1968). We refer to Ledwina and Wy{\l}upek (2012), p.~736, for a related discussion.

Proposition A implies that, under $F = G$ and arbitrary fixed $p \in (0,1)$, it holds that ${\sf A}_N(p) \stackrel{\cal D}{\longrightarrow} N(0,p(1-p))$ and ${\sf B}_N(p) \stackrel{\cal D}{\longrightarrow} N(0,p(1-p))$.  Hence, our estimates satisfy
$$
\eta_N \widehat{\text{CC}}(p) \stackrel{\cal D}{\longrightarrow} N(0,1),\;\;\;\mbox{and}\;\;\;\eta_N \widehat{\text{CCC}}(p) \stackrel{\cal D}{\longrightarrow} N(0,1).
\eqno(A5)
$$
This implies that, under $F = G$, in contrast to sample PP and QQ plots, given $p \in (0,1),$ the empirical comparison curves $\widehat{\text{CC}}(p)$ and $\widehat{\text{CCC}}(p)$ have (asymptotically) the same sensitivity to differences between $F$ and $G$ over the whole range of $p$'s. Note also that in the case $F = G$ asymptotic finite dimensional distributions of both processes ${\sf U}_N$ and  ${\sf P}_N$, pertinent to the respective comparison curves, are the same. 

In case when $F(x) \neq G(x)$, for some $x \in \mathbb R,$ the statements (A1), (A3) and (A2), (A4) exhibit how asymptotic variances and covariances of the processes ${\sf A}_N$ and ${\sf B}_N$ are influenced by the respective comparison densities. In particular, an impact of $r(p)$, in cases when it is unbounded near ends of (0,1), to the variance and covariance, (A1) and (A3), respectively,  can be noticeable. This drawback is obviously inherited by $\eta_N \widehat{\text{CC}}(p)$. These observations suggest that finite sample results for $\eta_N \widehat{\text{CCC}}(p)$ can be expected to be more stable than related results pertinent to $\eta_N \widehat{\text{CC}}(p)$.\\

\noindent
{\sf Some comments on derivation of (A1) and (A2)}\\

\noindent
In contrast to the source papers, in (A1) and (A2), for standardization we use a common in statistics term $\eta_N=\sqrt{mn/N}$. This implies related rescaling of the original expressions. With such rescaling, the results on ${\sf A}_N$ are immediate consequences of Theorem 3.1 in Aly et al. (1987).

In (A2), we rely on both empirical comparison distributions $F_m(H_N^{-1}(p))$ and  $G_n(H_N^{-1}(p))$. The results in Pyke and Shorack (1968) are phrased for $\sqrt N \Bigl[F_m(H_N^{-1}(p))-F(H^{-1}(p))\Bigr]$. However, it holds that  $F_m-G_n=(1-\lambda_N)^{-1}[F_m-H_{N}]$ and $F-G=(1-\lambda_N)^{-1}[F-H]$.  The above  yields
$$
\eta_N\Bigl\{[F_m(H_{N}^{-1}(p))-G_n(H_{N}^{-1}(p))]-[F(H^{-1}(p))-G(H^{-1}(p))] \Bigr\}=
$$
$$
\sqrt{\frac{\lambda_N}{1-\lambda_N}} \sqrt N  \Bigl\{[F_m(H_{N}^{-1}(p))-F(H^{-1}(p))] + [H(H^{-1}(p)) - H_N(H_N^{-1}(p))]\Bigr\}.
$$
Moreover, we have $\sup_{0 \leq p \leq 1}|H(H^{-1}(p)) - H_N(H_N^{-1}(p))| \leq 1/N$; cf. e.g. Pyke and Shorack (1968), p.~760.  Therefore, in view of the assumption on $\lambda_N$, the second term in the above formula is negligible and it is enough to use the existing results on the first term.  Hence, the conclusions follow.\\

\noindent
{\bf Appendix B. Some results on} $\;{\cal U}_{D(N)}$\\

\noindent
In view of consideration in Appendix E, we find it to be instructive to consider $\;{\cal U}_{D(N)}$ along with its `continuous' version
$$
{\cal U}^c_{\epsilon_N}=\inf_{\epsilon_N \leq p \leq 1-\epsilon_N} {\sf U}_N(p),\;\;\;\mbox{where}\;\;\epsilon_N \in (0,1).
$$
Under fixed $N$, theoretical derivations on  $\;{\cal U}_{D(N)}$ presented below are equivalent to that pertinent to  $\;{\cal U}^c_{\epsilon_N}$. In asymptotic results, to get the equivalence,  ranges of $D(N)$ and $\epsilon_N$ should be properly related.

To simplify reading of results stated below, for an event $A$, denote by $P(A;\bullet)$ a probability of $A$ calculated in the situation when the condition $\bullet$ is satisfied.\\

\noindent
{\bf Proposition B.1.} {\it For any $q \in \mathbb R$ and any $N \in \mathbb N$ it holds that}
$$
P({\cal U}_{D(N)} < q ; {\mathbb H}) \leq P({\cal U}_{D(N)} < q ; F=G)\;\;\;\mbox{and}\;\;\;P({\cal U}^c_{\epsilon_N} < q ; {\mathbb H}) \leq P({\cal U}^c_{\epsilon_N} < q ; F=G).
$$

\noindent
{\bf Corollary B.1.} {\it  For a significance level $\alpha \in (0,1)$ and sample sizes $m$ and $n$, introduce the largest value $q(\alpha;{\cal U}_{D(N)})$ such that 
	$P({\cal U}_{D(N)} < q(\alpha;{\cal U}_{D(N)}) ; F=G) \leq \alpha$. Then for all $F$ and $G$ from ${\mathbb H}$ we have $P({\cal U}_{D(N)} < q(\alpha;{\cal U}_{D(N)});{\mathbb H}) \leq \alpha$. Note also that, due to invariance of $G_n(F_m^{-1})$ to continuous strictly increasing transformations of the samples, the test statistic ${\cal U}_{D(N)}$ is distribution free under $F = G$. Similar conclusions hold  for $\;{\cal U}^c_{\epsilon_N}$ and $L({\sf{U}}_N,I_k)=\eta_N \min_{j \in I_k} \widehat{\text{CC}}(p_{S(N),j})$ defined in Section 5.2.} \\

\noindent
{\bf Remark B.1.} {\it  Test statistics $\;{\cal U}_{D(N)}$  and $\;{\cal U}^c_{\epsilon_N}$ have discrete finite sample distributions. We do not want to introduce randomized tests. In such situation one has to be careful with calculating quantiles, defining related critical values. This question is extensively discussed on p.~735 of Ledwina and Wy{\l}upek (2012). Following that contribution, we shall call $q(\alpha;\bullet )$ defined above  the critical value of $\bullet$, and calculate it by simulations, whenever needed, as described ibidem.}\\

\noindent
Asymptotic results  are stated below. Recall  that $0 < \lambda_* \leq \lambda_N \leq 1-\lambda_* < 1$ for some  $\;\lambda_* \leq 1/2$. 
Additionally  set $D^*(N)=\lfloor {N/\log^3N -1} \rfloor$, where $\lfloor \bullet \rfloor$ denotes the integer part of $\bullet$.\\

\noindent
{\bf Proposition B.2.} {\it Suppose that $F$ and $G$ have densities $f$ and $g$, such that $f(x) > 0$ on $\mathbb R$, while $g(x)$ is bounded on $\mathbb R$. Assume that  $D(N) \leq D^*(N)$ for $N$ sufficiently large. Then, for any such $F$ and $G$ from ${\mathbb A}$ and  for any $\alpha \in (0,1)$ it holds that
	$$
	\lim_{N \to \infty} P({\cal U}_{D(N)} < q(\alpha;{\cal U}_{D(N)}) ; {\mathbb A}) =1.
	$$
	Moreover, analogous  statement holds true for $\;{\cal U}^c_{\epsilon_N}$ provided that $\epsilon_N \geq \epsilon^*_N$ while $\epsilon^*_N=N^{-1} \log^3N$}.
\\

\noindent
{\bf Remark B.2.} {\it The assumption $D(N) \leq D^*(N)$, for sufficiently large $N$,  is stronger than related assumption $D(N)=o(N)$ required for consistency of $\;{\cal P}_{D(N)}$, when using a similar argument as in the proof of Lemma 2 (ii) in Ledwina and Wy{\l}upek (2012). The assumptions on $D(N)$'s are sufficient ones to get rates of growth of respective critical values. Though the assumptions are consequences of the techniques used to get related asymptotic results, finite sample experiments clearly show that, for large $N$, e.g.\ to obtain stable behavior of critical values of $\;{\cal U}_{D(N)}$ and pertinent barrier $l_{\sf U}(N,\alpha,I_1)$, one has to consider smaller $D(N)$ than applicable in the case of $\;{\cal P}_{D(N)}$ and pertinent barrier $l_{\sf P}(N,\alpha,I_1)$. See Section 5.4 and Appendix C for some evidence.
	
	Note also that, when one assumes that $D(N)$ and pertinent $S(N)$ are independent of $N$, consistency of $\;{\cal U}_{D(N)}$ takes place for all alternative models $F$ and $G$ such that $G(F^{-1}(p_{S(N),j})) - p_{S(N),j} < 0$ for some $j=1,...,S(N)$.}\\

\noindent
{\sf Proof of Proposition B.1} \\

\noindent
We apply  Lemma 5.9.1, p.~179, from Lehmann and Romano (2005). Similar reasoning in the case of $M_{D(N)}$, a forerunner of ${\cal P}_{D(N)}$, has already been presented in Appendix A.1 of Ledwina and Wy{\l}upek (2012).

Consider vectors  $(x_1,...,x_m,y_1,...y_n)$ and $(x_1,...,x_m,y^*_1,...y^*_n)$, where $y_i \leq y^*_i$ for each $i=1,...,n$. Let ${\tilde F}_m$, ${\tilde G}_n$ and ${\tilde G}^*_n$ stand for functions defined as related  empirical CDF's for components of these vectors of numbers. Note that ${\tilde G}_n \geq {\tilde G}^*_n$. Hence, given $p \in (0,1)$,  it holds that
$$
\frac{p- {\tilde G}_n\bigl({\tilde F}_m^{-1}(p)\bigr)}{\sqrt{p(1-p)}} \leq \frac{p- {\tilde G}^*_n\bigl({\tilde F}_m^{-1}(p)\bigr)}{\sqrt{p(1-p)}}.
$$
In consequence, the same relation holds for  minima of these expressions over the finite set of points, appearing in the related functions pertinent to  $\;{\cal U}_{D(N)}$. Recall also that $\;{\cal U}_{D(N)}$ rejects $\mathbb{H}$ for small observed values. Therefore, the assumption of Lemma 5.9.1 of Lehamnn and Romano (2005) is fulfilled.

Similar argument applies to  $\;{\cal U}^c_{\epsilon_N}$. \hfill{$\Box$}\\

\noindent
{\sf Proof of Proposition B.2} \\

\noindent 
For convenience, let us introduce an auxiliary statistic
$$
{\cal W}_{D(N)} = \max_{1 \leq j \leq D(N)}\{-{\sf U}_N(p_{S(N),j})\} = - {\cal U}_{D(N)}
$$
rejecting $\mathbb H$ for large values of ${\cal W}_{D(N)}$. Let $q(\alpha;{\cal W}_{D(N)})$ denote pertinent critical value.

We start with establishing a rate at which $q(\alpha;{\cal W}_{D(N)}) \to \infty$ under $F = G$. We have
$$
\max_{1 \leq j \leq D(N)}\{-{\sf U}_N(p_{S(N),j})\} \leq \max_{1 \leq j \leq D^*(N)}\{-{\sf U}_N(p_{S(N),j})\} \leq 
\sup_{\epsilon^*_N \leq p \leq {1-\epsilon^*_N}} \frac{\eta_{N} |p-G_n(F_m^{-1}(p))|}{\sqrt{p(1-p)}} \leq
$$
$$
\sup_{\epsilon^*_N \leq p \leq {1-\epsilon^*_N}} \frac{\eta_{N}|G(F_m^{-1}(p))-G_n(F_m^{-1}(p))|}{\sqrt{p(1-p)}}+
\sup_{\epsilon^*_N \leq p \leq {1-\epsilon^*_N}} \frac{\eta_{N}|p-G(F_m^{-1}(p))|}{\sqrt{p(1-p)}}.
\eqno(B1)
$$

Since ${\mathbb H}$ holds true, by Remark B.1, we can restrict attention to the case when $F(p)=G(p)=p$, for $p \in (0,1)$, and related uniform empirical and uniform quantile processes. Hence, Lemma 4.4.2 of Cs\"{o}rg\H{o} et al. (1986) applies and yields the rate $O_P(\sqrt{\log \log N})$ for the second component in (B1). To get a result for the first component in (B1), write
$$
\sup_{\epsilon^*_N \leq p \leq {1-\epsilon^*_N}} \frac{\eta_N|G(F_m^{-1}(p))-G_n(F_m^{-1}(p))|}{\sqrt{p(1-p)}} \leq 
{\frac{1}{\sqrt{{\epsilon^*_N}(1-\epsilon^*_N)}}} \times O_P(1)= O_P\Bigl(\sqrt{N/\log^3 N}\Bigr).
$$
Hence, $q(\alpha;{\cal U}_{D(N)})=O\Bigl(\sqrt{N/\log^3 N}\Bigr)$. The same relation holds for $q(\alpha;{\cal U}^c_{\epsilon_N})$ if for  the pertinent $\epsilon_N $ it holds that $\epsilon_N \geq \epsilon^*_N$. This implies that critical values of both statistics grow slower than $\sqrt N$.

Now we shall show that under ${\mathbb A}$ test statistics grow at the rate $\sqrt N$. In view of the above, this shall imply that $P({\cal U}_{D(N)} \leq q(\alpha;{\cal U}_{D(N)})) \to 1$ as $N \to \infty$. Similar relation holds for ${\cal U}^c_{\epsilon_N}$.
Under ${\mathbb A}$ and for sufficiently large $N$  there exists partition point $p_{s_0,j_0}$ such that CC$(p_{s_0,j_0})<0$. Also, for $N$ large enough it holds that $p_{s_0,j_0} \in [\epsilon^*_N, 1-\epsilon^*_N].$ Moreover, 
$P({\cal U}_{D(N)} < q(\alpha;{\cal U}_{D(N)})) \geq P({\sf U}_N (p_{s_0,j_0}) < q(\alpha;{\cal U}_{D(N)})).$ On the other hand,
$$
\sqrt N [p_{s_0,j_0}- G_n(F_m^{-1}(p_{s_0,j_0}))]=
\sqrt N [G(F_m^{-1}(p_{s_0,j_0}))-G_n(F_m^{-1}(p_{s_0,j_0}))]+
$$
$$
\sqrt N [G(F^{-1}(p_{s_0,j_0}))-G(F_m^{-1}(p_{s_0,j_0}))] +
\sqrt N [p_{s_0,j_0}-G(F^{-1}(p_{s_0,j_0}))].
\eqno(B2)
$$
The last component in (B2) is of the order $\sqrt N$. The first one behaves as classical Kolmogorov-Smirnov statistic and is of the order $O_P(1)$. In turn, for the middle term it holds that
$$
\sqrt N [G(F^{-1}(p_{s_0,j_0}))-G(F_m^{-1}(p_{s_0,j_0}))]=
$$
$$
\sqrt m [F^{-1}(p_{s_0,j_0})-F_m^{-1}(p_{s_0,j_0})]\times O(1) \times {g(\theta_m)},
$$
where $\theta_m$ is a random point lying in between $F^{-1}(p_{s_0,j_0})$ and $F_m^{-1}(p_{s_0,j_0})$. Since by assumption $g$ is bounded and $f$ is positive on $\mathbb R$,   then classical theorem on asymptotic normality of sample quantile implies that the middle term of (B2) is $O_P(1)$. The above concludes the proof. \hfill{$\Box$}\\

\noindent
{\bf Appendix C. Auxiliary simulation experiments}\\

\noindent
In this appendix, we present results of additional simulation experiments related to both income and cholesterol data analysis considered in Section 5.4, and contrast them with the previous ones. To be specific, we study the effect of choosing D(N) on corresponding critical values and acceptance regions. More precisely,  we present  results for $m$ and $n$ as pertinent to these data but with $D(N)$ being the largest natural number satisfying $D(N) \leq N$.

To  better see the differences resulting from passing from $D(N)=127$ to the above defined $D(N)$ we present for both processes ${\sf U}_N$ and ${\sf P}_N$ numerical values of pertinent simulated barriers $l_{\sf U}(N,\alpha,I_k)$ and $l_{\sf P}(N,\alpha,I_k)$ as well as obtained   numerical values of $L({\sf U}_N,I_k)$ and $L({\sf P}_N,I_k)$.

For completeness, recall that the critical values and barriers in acceptance regions are always simulated under $F=G$. Also, in both the barriers and the tests we take $\alpha=0.05$. Moreover, note that throughout the critical values, barriers in acceptance regions, and box plots are based on 100 000 Monte Carlo runs.\\

\noindent
{\sf C.1.  Canadian after-tax family income in 1978 versus 1986 comparison with D(N) = 16 383}\\[-1em]

\noindent
Here we consider the case $S(N)=13$ and  $D(N)=16\; 383$.  With such $S(N)$ some of corresponding $\;p_{S(N),j}$'s $\;$ falling into $I_1$ are much smaller than those appearing in $I_1$ under $S(N)=6$ and $D(N)=127$, considered in the main body of the paper. Tables C1, C2  and Figure C1, presented below, exhibit that in such extreme circumstances ${\cal P}_{D(N)}$ still shows stable behavior, while $\;{\cal U}_{D(N)}$ has a noticeably longer left tail. This leads to  considerable smaller critical values than obtained under $D(N)=127$. 
Also, related box plots indicate some instability in calculation of the barrier $L({\sf U}_N,I_1)$. \\

\begin{table}[h!]
	\label{Tab:C1}
	\renewcommand\thetable{C1} 
	\caption{Basic outcomes in  Canadian after-tax family income in 1978 versus 1986 comparison with $m=9\;470$ and $n=8\;526$. Results obtained for $D(N)=127$ and $D(N)=16\;383$.}
	
	\begin{center}	
		\begin{tabular}{l|cc|cc}
			\hline	

			                &\multicolumn{2}{c|}{$D(N)=127$}& \multicolumn{2}{c}{$D(N)=16\;383$}\\
			                \hline
			Test statistic	& $\mathcal{U}_{D(N)}$ & $\mathcal{P}_{D(N)}$& $\mathcal{U}_{D(N)}$ & $\mathcal{P}_{D(N)}$ \\
			\hline
			Obtained value	&-7.500  &-6.790 &-8.000 &-7.102 \\
			Critical value	&-2.854  &-2.848 &-4.027 &-3.140 \\
			p-value	        &0       &0 &0.001  &0	  \\
			\hline
		\end{tabular}
	\end{center}
\end{table}

\vspace*{-.5\baselineskip} 

\begin{figure}[h!]
	\centering
	\renewcommand\thefigure{C1} 
	\includegraphics[width=.75\linewidth]{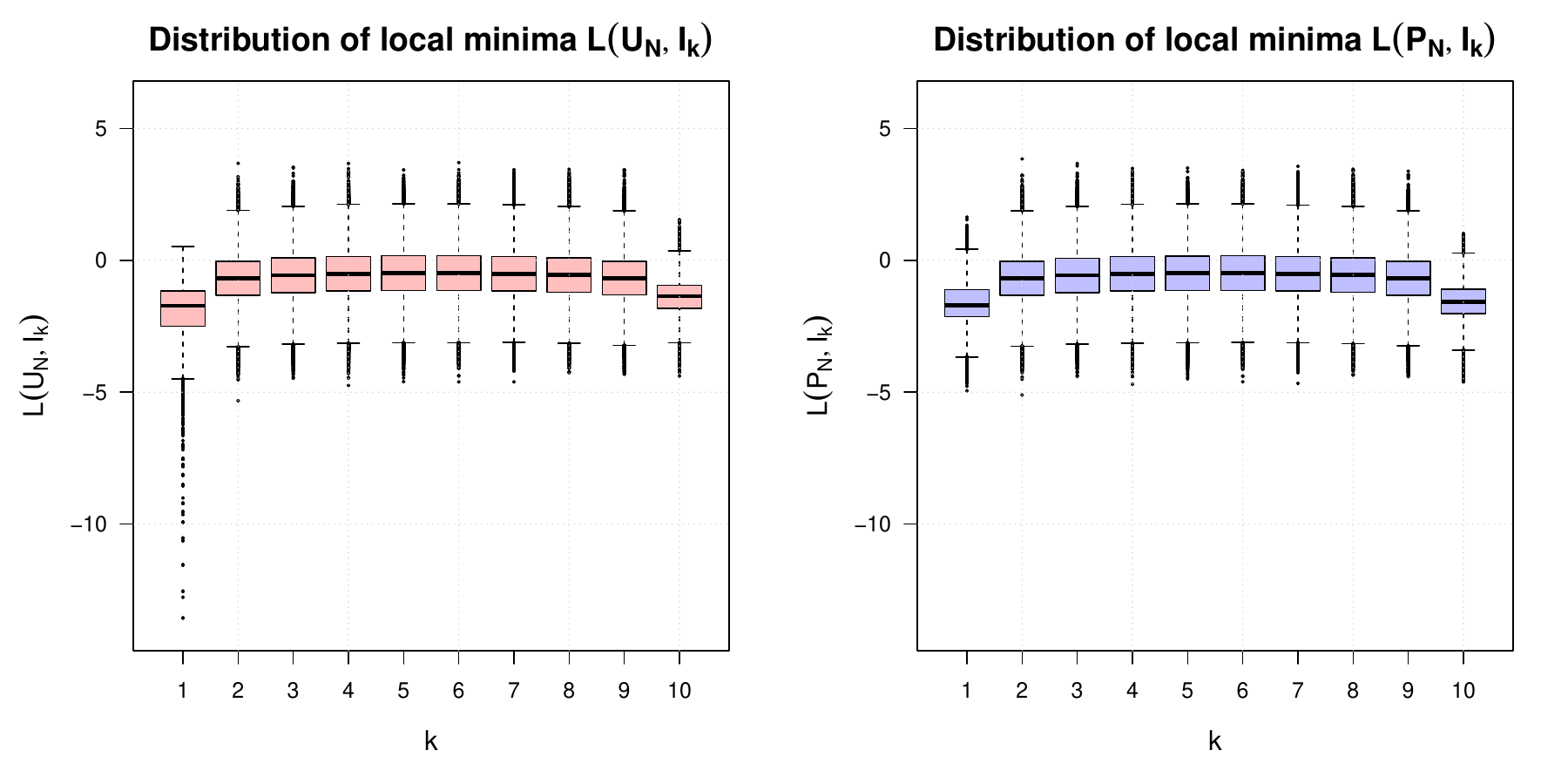}
	\caption{Results for  Canadian after-tax family income  in 1978 versus 1986 comparison with 
		$m = 9\; 470, n = 8\; 526$ and $D(N) = 16\; 383$. The figure shows
		box plots obtained for distributions of the barriers  $L({\sf U}_N,I_k)$ and $L({\sf P}_N,I_k),\; k=1,...,10$, under $F = G$.}		
	\label{Fig:C1}
\end{figure}

To gain a deeper insight into the behaviour of processes  ${\sf U}_N(p)$ and ${\sf P}_N(p)$,  in Table C2 we compare numerical values of simulated barriers in corresponding acceptance regions, i.e. $l_{\sf U}(N,\alpha,I_k)$ and $l_{\sf P}(N,\alpha,I_k)$, as well as observed values of $L({\sf U}_N,I_k)$ and $L({\sf P}_N,I_k)$. Results were  obtained under $D(N)=127$ and $D(N)=16\;383$, respectively.  
We see that in case of $D(N)= 16\; 383$, an instability of unpooled approach for $p \in(0,0.1]$ is considerable, and it leads to significantly smaller  value of the corresponding barrier $l_{\sf U}(N,\alpha,I_1)$ comparing to $D(N)=127$.
On the other hand, pooled approach does not suffer from such drawback and ${\cal P}_{D(N)}$ reacts very moderately to the transition from $D(N)=127$ to $D(N)=16\; 383$.\\


\begin{table}[h!]
	\renewcommand\thetable{C2}
	\caption{Numerical values of simulated barriers $l_U(N,\alpha,I_k)$ and $l_P(N,\alpha,I_k)$ in acceptance regions
		as well as obtained values of $L({\sf U}_N,I_k)$ and $L({\sf P}_N,I_k)$ for 1978 versus 1986 after-tax income comparison. The values lying below pertinent barriers are marked in boldface.
		Results obtained for $D(N)= 127$ and $D(N)=16\;383$.}
\begin{center}
{\small
		\begin{tabular}{l|cccccccccc}
			\hline
			& \multicolumn{10}{|c}{D(N) = 127}\\
			 \cline{2-11}
			& $I_1$ &$I_2$ &$I_3$ &$I_4$ &$I_5$ &$I_6$ &$I_7$ &$I_8$ &$I_9$ &$I_{10}$ \\
			\hline
			$l_{\sf U}(N,\alpha,I_k)$ & -2.393 &-2.135 &-2.052 &-2.018 &-1.995 &-1.971 &-1.999 &-2.028 &-2.081 &-2.224 \\
			$L({\sf U}_N,I_k)$ & {\bf-7.500} & 0.310 & 1.579 & 2.552 & 1.791 & 0.251 & {\bf -2.351} &{\bf -4.449 }&{\bf -4.523} &{\bf-3.809}\\
			\hline
			$l_{\sf P}(N,\alpha,I_k)$ &  -2.335 &-2.103 &-2.037 &-2.006 &-1.987 &-1.983 &-2.010 &-2.045 &-2.105 &-2.303\\
			$L({\sf P}_N,I_k)$ &{\bf -6.790} & 0.323 & 1.674 & 2.568 & 1.656 & 0.331 &-1.806 &{\bf -4.365} &{\bf -4.655} &{\bf -3.966}\\
			\hline
			& \multicolumn{10}{|c}{D(N) = 16\;383}\\
			 \cline{2-11}
			& $I_1$ &$I_2$ &$I_3$ &$I_4$ &$I_5$ &$I_6$ &$I_7$ &$I_8$ &$I_9$ &$I_{10}$ \\
			\hline
			$l_{\sf U}(N,\alpha,I_k)$ &-3.982 &-2.279 &-2.179 &-2.128 &-2.094 &-2.077 &-2.096 &-2.126 &-2.215 &-2.495 \\
			$L({\sf U}_N,I_k)$ &{\bf -8.000} & 0.129 &1.547 &2.398 &1.663 &0.218 &{\bf-2.469} &{\bf -4.658} &{\bf -4.664} &{\bf -4.005}\\
			\hline
			$l_{\sf P}(N,\alpha,I_k)$ &-2.817 &-2.252 &-2.165 &-2.122 &-2.097 &-2.084 &-2.107 &-2.146 &-2.251 &-2.755\\
			$L({\sf P}_N,I_k)$ &{\bf -7.102}  &0.133  &1.604 &2.405 &1.586 &0.212 &{\bf -2.193} &{\bf -4.614} &{\bf -4.794} &{\bf -4.218}\\
			\hline
		\end{tabular}
}
\end{center}
\end{table}



\noindent
{\sf C.2. Cholesterol level in obese men in Puerto Rico versus Honolulu  comparison with D(N) = 511}\\[-1em]

\noindent
In this section we present results obtained in the case $S(N)=8$ and  $D(N)=511$, i.e. the largest natural number satisfying $D(N) \leq N$.
Recall that this case differs from the previous example in that the sample sizes are much smaller and evidently 
unbalanced ones.
 Tables C3, C4, and Figure C2, presented below, indicate that ${\cal P}_{D(N)}$ still shows stable behavior, while $\;{\cal U}_{D(N)}$ exhibits in $I_1$ even larger instability than in the case $S(N)=6$ and $D(N)=127$, considered in the main body of the paper. 
Again, as in the previous example, this leads to considerable smaller critical values than obtained under $D(N)=127$.  Moreover, related box plots confirms instability in the calculation of the barrier $L({\sf U}_N,I_1)$.


\begin{table}[h!]
	\renewcommand\thetable{C3}
	\caption{Basic outcomes in Puerto Rico versus Honolulu comparison of cholesterol levels in obese men groups with $m=160$ and $n=628$.
    Results obtained for $D(N)=127$ and $D(N)=511$.}
	\label{Tab:C3}
	\begin{center}	
		\begin{tabular}{l|cc|cc}
			\hline
	                & \multicolumn{2}{c|}{$D(N)=127$} & \multicolumn{2}{c}{$D(N)=511$}\\
	                \hline
	Test statistic	& $\mathcal{U}_{D(N)}$ & $\mathcal{P}_{D(N)}$& $\mathcal{U}_{D(N)}$ & $\mathcal{P}_{D(N)}$ \\
			\hline
			Obtained value	&-6.019 &-5.038 & -6.123 &-5.045 \\
			Critical value	&-3.491 &-2.813 &-4.795  &-2.982 \\
			p-value	   &0.003 &0  &0.021 &0  \\
			\hline
		\end{tabular}
	\end{center}
\end{table}


\begin{figure}[h!]
	\renewcommand\thefigure{C2}
	\centering
	\includegraphics[width=.75\linewidth]{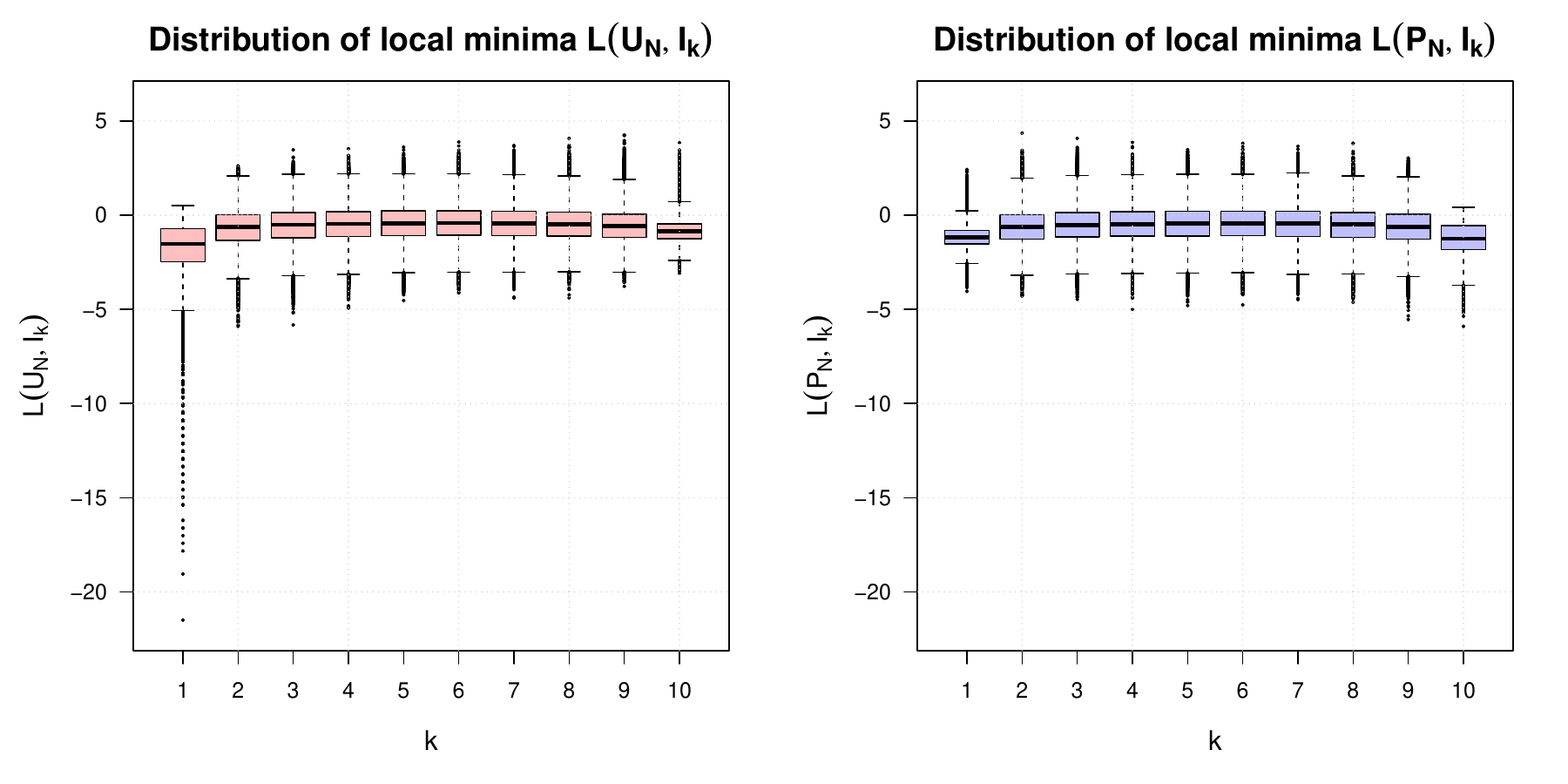}
	\caption{Results for comparison of Puerto Rico versus Honolulu cholesterol levels in obese men groups with
		$m=160, n=628$ and $D(N)=511$. The figure shows box plots obtained for distributions of the barriers  $L({\sf U}_N,I_k)$ and $L({\sf P}_N,I_k),\; k=1,...,10$, under $F = G$.}		
	\label{Fig:C2}
\end{figure}

Similarly as in Appendix C1, in Table C4 we compare numerical values of simulated barriers in corresponding acceptance regions as well as observed values of $L({\sf U}_N,I_k)$ and $L({\sf P}_N,I_k)$.
Again, unpooled approach ${\sf U}_{N}(p)$ exhibits significant instability  in $I_1$, while ${\sf P}_{N}(p)$ responds very moderately to the change from  $D(N)=127$ to $D(N)=511$. 

\begin{table}[h!]
	\renewcommand\thetable{C4}
	\caption{Numerical values of simulated barriers $l_U(N,\alpha,I_k)$ and $l_P(N,\alpha,I_k)$ in acceptance regions
		as well as obtained values of $L({\sf U}_N,I_k)$ and $L({\sf P}_N,I_k)$ for comparison of Puerto Rico versus Honolulu cholesterol levels in obese men groups.	The values lying below pertinent barriers are marked in boldface.	Results obtained for $D(N)= 127$ and $D(N)=511$.}
	\begin{center}
	\label{Tab:C4}	
{\small
		\begin{tabular}{l|cccccccccc}
			\hline
			& \multicolumn{10}{|c}{D(N) = 127}\\
			\cline{2-11}
			& $I_1$ &$I_2$ &$I_3$ &$I_4$ &$I_5$ &$I_6$ &$I_7$ &$I_8$ &$I_9$ &$I_{10}$ \\
			\hline
			$l_{\sf U}(N,\alpha,I_k)$ & -3.365 & -2.335 & -2.141 & -2.036 & -1.972 &-1.927 &-1.928 & -1.885 & -1.832 & -1.679 \\ 
			$L({\sf U}_N,I_k)$ &{\bf -4.513} &{\bf -6.019} &{\bf-5.841} &{\bf-4.792} &{\bf-3.913} &{\bf-3.677} &{\bf-2.767} &{\bf-2.379} &-1.118 &-0.698\\
			\hline
			$l_{\sf P}(N,\alpha,I_k)$ & -2.075 & -2.054 & -2.011 & -1.968 & -1.975 & -1.988 & -2.002 & -2.072 & -2.161 & -2.545\\ 
			$L({\sf P}_N,I_k)$ &{\bf-2.685} &{\bf-3.453} &{\bf-4.434} &{\bf-4.833} &{\bf-5.038} &{\bf-4.125} &{\bf-3.699} &{\bf-2.912} &-1.623 &-0.402\\
			\hline
			& \multicolumn{10}{|c}{D(N) = 511}\\
			\cline{2-11}
			& $I_1$ &$I_2$ &$I_3$ &$I_4$ &$I_5$ &$I_6$ &$I_7$ &$I_8$ &$I_9$ &$I_{10}$ \\
			\hline
			$l_{\sf U}(N,\alpha,I_k)$ & -4.795 &-2.419 &-2.216 &-2.121 &-2.057 &-2.007 &-1.978 &-1.953 &-1.931 &-1.776 \\
			$L({\sf U}_N,I_k)$ &{\bf -5.012} &{\bf-6.123} &{\bf-5.990} &{\bf-4.905} &{\bf-4.057} &{\bf-3.720} &{\bf-2.895} &{\bf-2.422} &-1.220 &-0.826\\
		\hline
			$l_{\sf P}(N,\alpha,I_k)$  &-2.190 &-2.146 &-2.070 &-2.031 &-2.049 &-2.059 &-2.101 &-2.154 &-2.274&-2.751\\
			$L({\sf P}_N,I_k)$ &{\bf-2.685} &{\bf-3.453} &{\bf-4.444} &{\bf-4.852} &{\bf-5.045} &{\bf-4.126} &{\bf-3.707} &{\bf-2.912} &-1.646 &-0.965\\
			\hline
		\end{tabular}
}
	\end{center}
\end{table}


\clearpage

\noindent
{\bf Appendix D. Some comments on relation of CC  and CCC}\\

\noindent
There is a close relationship between pooled and unpooled relative distributions. Recall that we are focusing on continuous $F$ and $G$. Therefore, we infer  that
$$
H(F^{-1}(p))=\lambda_N p + (1-\lambda_N) G(F^{-1}(p)),\;\;\;\mbox{and}\;\;\;\Bigl[H\circ F^{-1}\Bigr]^{-1}(p)=F(H^{-1}(p)) \;\;\;p \in [0,1].
$$
Hence, $F(H^{-1}(p))$ is the inverse of $\lambda_N p + (1-\lambda_N) G(F^{-1}(p))$. This yields
$$
r_1(p)=\frac{dF(H^{-1}(p))}{dp} =\frac{1}{\lambda_N + (1-\lambda_N) r(F(H^{-1}(p))}, 
$$
with $r(p)=dG(F^{-1}(p))/dp$. Thus, it is seen that comparison densities in the pooled and unpooled settings are explicitly related.

CC and CCC curves provide two ways of inter-distributional comparison. Often their shapes are similar. But, obviously, one can imagine some pairs of $(F,G)$ for which visible differences between the two curves shall appear. In particular, it should be noted that CCC takes into account available sample sizes $m$ and $n$, while CC ignores this information. Introducing more precise notation $\text{CC}(p)=\text{CC}_{(G,F)}(p)=[p-G(F^{-1}(p))]/\sqrt{p(1-p)}$, we see that
$\text{CCC}(p)=\text{CC}_{(F,H)}(p)-\text{CC}_{(G,H)}(p)$. Hence, for any fixed $p \in (0,1)$, it holds that $\text{CCC}(p) \to - \text{CC}_{(F,G)}(p)$ if $\lambda_N \to 0$, and $\text{CCC}(p) \to \text{CC}_{(G,F)}(p)$ if $\lambda_N \to 1$. See Sections 5.3 and 5.4 of our paper for some illustrations. 

Section 2.4.1 of Handcock and Morris (1999) contains some remarks contrasting pooled and unpooled approaches. Their book focuses on the unpooled approach and their choice of the reference group is motivated by an easier interpretability of unpooled case over the pooled one. Also, in their inference, they mostly rely on pertinent relative densities and their estimates. We prefer relative CDF's, their standardized versions, and related empirical variants, as on the level on densities not all differences between two populations are well seen. It seems that by introducing a contrast comparison curve, we have removed  objection concerning interpretation of outcomes of pooled approach. Also, grater stability of empirical CCC over CC, exemplified in this paper, encourages to consider pooled approach, and CCC as a useful competitor of CC.\\

\pagebreak
\noindent{\bf Appendix E. New summary indices and related discussion}\\

\noindent
A number of indices, summarizing an information contained in the ROC curve have been introduced. For example, Greenhouse and Mantel (1950) proposed ROC($p_0$), for a fixed specificity $1- p_0$. Several researches have used the area under ROC curve (AUC); see Hilden (1991) and references therein. Wieand et al. (1989) proposed a weighted area under ROC, while McClish (1989) advocated for a  partial area under the curve. Vexler at al. (2018, 2019) 
	discussed some applications of AUC based approaches to construction of some new inferential procedures.
For more indices and more detailed information on them see Krzanowski and Hand (2009), and \'Alvarez-Esteban et al. (2017).

Parallel studies in econometrics, generally completely independent from those in the field of biometry, have introduced several notions of restricted stochastic dominance and related indices. For an overview see Kamihigashi and Stachurski (2014a,b). In particular, they extensively discussed the index $\sigma(F,G)=1-\sup_{x \in {\mathbb R}}\{G(x)-F(x)\}$, measuring the extent to which $F$ is dominated by $G$.

At least part of empirical variants of the above mentioned summary indices define test statistics for the testing problem  
$$
{\mathbb H}^0 : F(x)=G(x),\;\;\;\mbox{for all}\;\;\;x \in \mathbb R,
$$
against unrestricted alternative
$$
{\mathbb A}^0 : F(x) \neq G(x),\;\;\; \mbox{for some}\;\;\;x \in \mathbb R.
$$ 

Looking from the point of view of testing ${\mathbb H}^0$,  one can  suggest a simple index $\sup_{x \in {\mathbb R}}|F(x)-G(x)|$. In fact, this quantity coincides with the well known Youden (1950) index, introduced in the context of searching for an optimal classification threshold. For related discussion see Chapters 2 and 4 in Krzanowski and Hand (2009). 

In general, to summarize an information contained in ROC curve, some suprema of adequate weighted variants of pertinent curves can be more reliable. Having introduced CC$(p)$ and CCC$(p)$,  the following indices can be immediately proposed
$$
\sup_{\epsilon \leq p \leq 1-\epsilon} |\text{CC}(p)|\;\;\;\mbox{and}\;\;\;\sup_{\epsilon \leq p \leq 1-\epsilon} |\text{CCC}(p)|,\;\;\;\epsilon \in (0,1),
\eqno(E1)
$$
and can be considered as a kind of counterparts of the partial area under the curve index. Since CC$(p)$ and CCC$(p)$ can be unbounded near 0 and 1, therefore $\epsilon=0$ is not a good choice, in general. Their empirical counterparts take the form
$$
\sup_{\epsilon \leq p \leq 1-\epsilon} \eta_N \times |\widehat{\text{CC}}(p)|\;\;\;\mbox{and}\;\;\;\sup_{\epsilon \leq p \leq 1-\epsilon} \eta_N \times  |\widehat{\text{CCC}}(p)|,\;\;\;\epsilon \in (0,1),
\eqno(E2)
$$

The second statistic in (E2) is expected to be more stable than the first one. The first statistic in (E2) resembles to some extent maximally selected chi-square statistic, designed for the pooled approach, in Miller and Siegmund (1982).  Moreover, this statistic  is similar in structure to the  statistic ${\cal E}_n$, introduced and studied in \'Cmiel et al. (2020), and designed for testing simple goodness of fit hypothesis. 
For related discussion on some weighted statistics for ${\mathbb H}^0$ see Doksum and Sievers (1976), Thas (2010),  Ledwina and Wy{\l}upek (2012) and references therein. 

The last comment concerns the range $[\epsilon,1-\epsilon]$ appearing in (E2).  Formally, we could insert the whole (0,1) interval here, as it takes place in the classical sup-type Anderson-Darling statistic.  Some  drawbacks of the last mentioned statistic are discussed in Section 2.13 of Milbrodt and Strasser (1990). Also, as shown recently in \'Cmiel et al. (2020), such solution is not a well balanced omnibus test. A modification of Anderson-Darling solution using Erd\"{o}s-Feller-Kolmogorov-Petrovski-type weight function removes some of the drawbacks but does not improve essentially power behavior of such sup-type statistic under moderate sample sizes; cf. \'Cmiel et al. (2020) for some evidence. Therefore, we do prefer to restrict slightly the range of $p$'s.  Evidently, for testing ${\mathbb H}_0$ one can consider more flexible solution with $\epsilon=\epsilon_N$. For such case, the barrier $\epsilon^*_N$ specified in Proposition B.2 of Appendix B is the minimal value, for which some useful limiting results hold.

Obviously, many other statistical functionals of the empirical processes ${\sf U}_N(p)$ and ${\sf P}_N(p)$, $p \in (0,1)$, including some standards as e.g. ROC based integral Anderson-Darling statistic, can be constructed. A need to base some indices summarizing ROC on empirical ROC function has been emphasized recently in Pardo and Franco-Pereira (2017). \\

\noindent
{\bf References for Appendices}\\

\setlist{nolistsep} 
\begin{spacing}{1}  
\begin{description}[topsep=0pt,itemsep=0pt,parsep=0pt,labelsep=0em]
	\item Aly, E.-E.A.A., Cs\"{o}rg\H{o}, M., and Horv\'ath, L. (1987). P-P plots, rank processes, and Chernoff-Savage theorems. In: {\it New Perspectives in Theoretical and Applied Statistics}, M.L. Puri, J.P. Vilaplana, and W. Wertz, eds., Wiley, New York, pp. 135-156.
	\item \'Alvarez-Esteban, P.C., del Barrio, E., Cuesta-Albertos, J.A., and Matr\'an (2017). Models for the assessment of treatment improvement: The ideal and the feasible. {\it Statistical Science} {\bf 32}, 469-485.
	\item Beirlant, J., and Deheuvels, P. (1990). On the approximation of P-P and Q-Q plot processes by Brownian bridges. {\it Statistics $\&$ Probability Letters} {\bf 9}, 241-251.
	\item Cs\"{o}rg\H{o}, M., Cs\"{o}rg\H{o}, S., Horv\'ath, L., and Mason, D.M. (1986). Weighted empirical and quantile processes. {\it Annals of  Probability} {\bf 14}, 31-85.
	\item \'Cmiel, B., Inglot, T., and Ledwina, T. (2020). Intermediate efficiency of some weighted goodness-of-fit statistics. {\it Journal of Nonparametric Statistics}, {\bf 32} 667-703.
	\item Doksum, K.A., and Sievers, G.L. (1976). Plotting with confidence: Graphical comparisons of two populations. {\it Biometrika} {\bf 63}, 421-434.
	\item Greenhouse, S.W., and Mantel, N. (1950). The evaluation of diagnostic tests. {\it Biometrics} {\bf 6}, 399-412.
	\item Handcock, M.S., and Morris, M. (1999). {\it Relative Distribution Methods in the Social Sciences}, Springer, New York.
	\item Hilden, J. (1991). The area under the ROC curve and its competitors. {\it Medical Decision Making} {\bf 11}, 95-101. 
	\item Kamihigashi, T., and Stachurski, J. (2014a). An axiomatic approach to measuring degree of stochastic dominance. {\it Discussion Paper DP2014-36}, Kobe University, Japan. 
	\item Kamihigashi, T., and Stachurski, J. (2014b). Partial stochastic dominance. {\it Discussion Paper DP2014-23}, Kobe University, Japan. 
	\item Krzanowski, W.J., and Hand, D.J. (2009). {\it ROC Curves for Continuous Data}, Chapman and Hall, London.
	\item Ledwina, T., and Wy{\l}upek, G. (2012). Nonparametric tests for first order stochastic dominance. {\it Test} {\bf 21}, 730-756.
	\item Lehmann, E.L., and Romano, J.P. (2005). {\it Testing Statistical Hypotheses}, Springer, New York.
	\item McClish, D. (1989). Analyzing a portion of the ROC curve. {\it Medical Decision Making} {\bf 9}, 190-195.
	\item Milbrodt, H., and Strasser, H. (1990). On the asymptotic power of the two-sided Kolmogorov-Smirnov test. {\it Journal of Statistical Planning and Inference} {\bf 26}, 1-23.
	\item Miller, R., Siegmund, D. (1982). Maximally selected chi square statistics. {\it Biometrics} {\bf 38}, 1011-1016.
	\item Pardo, M.C., and Franco-Pereira, A.M. (2017). Non parametric ROC summary statistics. {\it REVSTAT - Statistical Journal} {\bf 15}, 583-600.
	\item Pyke, R. (1970). Asymptotic results for rank statistics. In: {\it Nonparametric Techniques in Statistical Inference}, M.L. Puri, ed., Cambridge University Press, London, pp. 21-37.
	\item Pyke, R., and Shorack, G.R. (1968). Weak convergence of a two-sample empirical process and a new approach to Chernoff-Savage theorems. {\it Annals of Mathematical Statistics} {\bf 39}, 755-771.
	\item Thas, O. (2010). {\it Comparing Distributions}, Springer, New York.
	\item Vexler, A., Yu, J., Zhao, Y., Hutson, A. D., and Gurevich, G. (2018). Expected p-values in light of an ROC curve analysis applied to optimal multiple testing procedures. {\it Statistical Methods in Medical Research} {\bf 27}, 3560-3576.
	\item Vexler, A., Afendras, G., and Markatou, M. (2019). Multi-panel Kendall plot in light of an ROC curve analysis applied to measuring dependence. {\it Statistics} {\bf 53}, 417-439.
	\item Wieand, S., Gail, M.H., James, B.R., and James, K.L. (1989). A family of nonparametric statistics for comparing diagnostic markers with paired and unpaired data. {\it Biometrika} {\bf 76}, 585-592.
	\item Youden, W.J. (1950). Index for rating diagnostic tests. {\it Cancer} {\bf 3}, 32-35.
\end{description}
\end{spacing}

\end{document}